\documentclass[twocolumn]{aastex7}
\usepackage{amsmath}
\usepackage{wasysym}      
\usepackage{graphicx}
\usepackage{amssymb}
\usepackage{epstopdf}
\usepackage{mathrsfs}
\usepackage{anyfontsize}
\usepackage{natbib}
\usepackage{color}
\usepackage{lipsum}
\usepackage{diagbox}

\newcommand{\Msun}{M_{\odot}}

\shorttitle{The Peak of the Fallback Rate from TDEs}
\begin{document}
\title{Predicting the Properties of the Fallback Rate from Tidal Disruption Events: Investigating the Maximum Gravity Model}

\author[0009-0008-5847-9778]{Julia Fancher}
\affiliation{Department of Physics, Syracuse University, Syracuse, NY 13210, USA}
\email[show]{jlfanche@syr.edu}

\author[0000-0002-5116-844X]{Ananya Bandopadhyay}
\affiliation{Department of Physics, Syracuse University, Syracuse, NY 13210, USA}
\email[show]{abandopa@syr.edu}

\author[0000-0003-3765-6401]{Eric R.~Coughlin}
\affiliation{Department of Physics, Syracuse University, Syracuse, NY 13210, USA}
\email{ecoughli@syr.edu}

\author[0000-0002-2137-4146]{C.~J.~Nixon}
\affiliation{School of Physics and Astronomy, Sir William Henry Bragg Building, Woodhouse Ln., University of Leeds, Leeds LS2 9JT, UK}
\email{c.j.nixon@leeds.ac.uk}

\begin{abstract}
A star destroyed by the tidal field of a supermassive black hole (SMBH) in a tidal disruption event (TDE) gives rise to a luminous flare. TDEs are being detected at an ever-increasing rate, motivating the need for accurate models of their lightcurves. The ``maximum gravity'' (MG) model posits that a star is completely destroyed when the tidal field of the SMBH exceeds the maximum self-gravitational field within the star, $g_{\rm max}$, and predicts the peak fallback rate $\dot{M}_{\rm peak}$ and the time to peak $t_{\rm peak}$. Here we perform hydrodynamical simulations of the complete disruption of 24 stars with masses ranging from $0.2-5.0 M_\odot$, at different stages of their main sequence evolution, to test the predictions of this model. We find excellent agreement between the MG model predictions and our simulations for stars near the zero-age main sequence, while the predictions are less accurate (but still within $\sim 35-50\%$ of the simulation results) for highly evolved stars. We also generalize the MG model to incorporate the Paczy{\'n}ski-Wiita potential to assess the impact of strong-gravity effects -- which are especially important for deep encounters that are required to completely destroy evolved and centrally concentrated stars -- and find good agreement with recent works that include relativistic gravity. Our results demonstrate that this model provides accurate constraints on the peak timescale of TDE lightcurves and their correlation with black hole mass. 
\end{abstract}

\keywords{Astrophysical black holes (98); Supermassive black holes (1663); Black hole physics (159); Hydrodynamics (1963); Tidal disruption (1696)}

\section{Introduction}
A tidal disruption event (TDE) occurs when a star is partially or completely destroyed by the tidal field of a supermassive black hole (SMBH) \citep{hills75,rees88,gezari21}. Approximately half of the tidally stripped debris is bound to the black hole and circularizes to form an accretion disk \citep{cannizzo90, loeb97, hayasaki13, coughlin14, shiokawa15,bonnerot16,s16,curd19,andalman22, steinberg24, meza25}. The process of accretion releases energy at a rate comparable to or in excess of the Eddington limit of the SMBH \citep{evans89,wu18}, seen in the form of a luminous electromagnetic flare with -- at least in some cases -- associated outflows \citep{bloom11, cenko12, miller15, brown15, alexander16, pasham18, andreoni22, pasham23b, pasham23, wevers24, ajay25}. The detection rate of these flares has increased steadily in the past years 
(\citealt{arcavi14,french16,nicholl19,pasham19,wevers19,hinkle21,payne21,vanvelzen21,wevers21,lin22,nicholl22,hammerstein23,wevers23,yao23,guolo24,ho25}; see also \cite{gezari21} and references therein), and with the advent of the Vera Rubin observatory \citep{ivezic19} will increase further.

The fallback rate $\dot{M}$, or the rate at which the stellar material returns to the SMBH, is the main factor in determining the properties of the accretion disk in the first few years of disk formation \citep{rees88, cannizzo90, lodato11}. Analytic models of the fallback rate of a TDE are thus useful in characterizing these systems, the earliest example of which is known as the frozen-approximation \citep{lacy82,carter82, bicknell83,stone13}. This model assumes that the star is completely and instantaneously destroyed as it passes through the sphere of radius $r_{\rm{t}}$ from the SMBH, where $r_{\rm t}$ -- the tidal radius -- is determined by equating the surface gravitational field of the star with the tidal acceleration due to the SMBH, which yields~\citep{hills75}
\begin{equation}
    r_{\rm{t}} = R_\star \left( \frac{M_\bullet}{M_\star}\right)^{1/3}. \label{eq:rt}
\end{equation}
Here $M_\bullet$ is the mass of the SMBH and $M_\star$ and $R_\star$ are the stellar mass and radius, respectively. In this model, the peak of the fallback rate and the time to peak (see Figures \ref{fig:zamsplot} and \ref{fig:mams+tamsfallbackrates} below for examples) are determined by the bulk properties of the star and the canonical tidal radius $r_{\rm t}$ (as defined in Equation~\ref{eq:rt}), and are strongly dependent on stellar type (see \citealt{lkp}, and in particular their generalization to include the stellar density profile; see also \citealt{gallegos-garcia18}). While this model's predictions are in reasonable agreement with numerical simulations of the disruption of low-mass and polytropic stars \citep{lacy82,evans89,lkp}, the predictions for higher-mass stars are highly discrepant (and even inverted, i.e., the model predicts that the fallback rate for the disruption of a solar-like star peaks at a later time and at smaller magnitude compared to a low-mass star, while the opposite is observed in simulations; \citealt{golightly19}), and the strong dependence of the fallback rate on stellar properties predicted by this model is generally not found \citep{bandopadhyay24}. 

These discrepancies have been noted by, e.g., \citet{guillochon13, goicovic19, golightly19,lawsmith19,lawsmith20,ryu20a,jankovic23}. Some of these investigations generated empirical (i.e., based on the results of numerical experiments) scalings for the impact parameter $\beta\equiv r_{\rm t}/r_{\rm p}$ (where $r_{\rm p}$ is the pericenter distance) at which a star is completely destroyed, denoted $\beta_{\rm c}$, and the corresponding distance $r_{\rm t, c}$. For example,~\cite{lawsmith20} found that the critical impact parameter scales as $\beta_{\rm c} \sim 0.5 \, (\rho_{\rm c}/\rho_{\star})^{1/3}$, where $\rho_{\rm c}$ ($\rho_{\star}$) is the central (average) stellar density, for stars having $\rho_{\rm c}/\rho_{\star} \lesssim500$, while $\beta_{\rm c} \sim 0.39 \, (\rho_{\rm c}/\rho_{\star})^{1/2.3}$ when $\rho_{\rm c}/\rho_{\star} \gtrsim500$. \citet{ryu20b} also found that $\beta_{\rm c} \propto \left(\rho_{\rm c}/\rho_{\star}\right)^{1/3}$. 

These scalings were based on simulations of TDEs. In contrast,~\cite{coughlin22a} proposed that a star is completely destroyed if the tidal field of the SMBH exceeds the maximum self-gravitational field that occurs at some nonzero radius in the stellar interior (denoted the ``core radius,'' $R_{\rm c}$), which is always greater than the surface gravity. Using this definition of the core radius, they predicted the distance from the SMBH at which a star is completely destroyed, and derived analytical expressions for the peak fallback rate $\dot{M}_{\rm peak}$ and the time at which $\dot{M}_{\rm peak}$ is reached, $t_{\rm peak}$  
(which can be evaluated for any stellar density profile). This model, which we refer to as the maximum gravity (MG) model, does not make any approximations specific to stellar type, e.g., based on the ratio of $\rho_{\rm c}/\rho_{\star}$. Nonetheless, they showed that if one extrapolates the $\propto r$ behavior of the self-gravitational field from the stellar center and the $\propto 1/r^2$ behavior from the surface, the radius at which the two intersect yields an approximation for the core radius and 
the aforementioned and empirically obtained scaling $\beta_{\rm c} \propto (\rho_{\rm c}/\rho_{\star})^{1/3}$.

While \citet{coughlin22a} noted the accuracy of this model when compared to extant numerical results and \citet{bandopadhyay24} directly tested it against a handful of hydrodynamical simulations, it has not been vetted over a wide range of stellar progenitors. The aims of this work are: 1) to analyze the accuracy of the MG model when compared to numerical hydrodynamical simulations of the disruption of a wide range of stars in terms of their mass and age; 2) assess whether the approximate solution for $r_{\rm t,c}$ coupled with the expression for $t_{\rm peak}$ yields good agreement with the numerically obtained values, and if a better estimate can be obtained by using higher order approximations for the self-gravitational field; and 3) extend the MG model to include relativistic effects. In Section \ref{sec:MG-model} we provide a brief overview of the MG model, outline its predictions for the peak fallback rate from TDEs, and test the predictions of the MG model against numerical hydrodynamical simulations. We find excellent agreement between the MG model predictions and the numerical results for stars near the zero-age main sequence (ZAMS), while more highly evolved and centrally concentrated stars are not completely destroyed at the $\beta_{\rm c}$ predicted by the model. The fallback rates from these disruptions peak at later times and with smaller magnitudes than predicted, but are still generally within $\sim few\times 10\%$ of the MG values. In Section \ref{sec:pw-potential} we use the Paczy{\'n}ski-Wiita potential (in place of the Newtonian potential) in the MG model to estimate the impact of relativistic effects, and find -- despite the simplicity of this approximation -- good agreement with other works. We discuss the implications of our results and conclude in Section \ref{sec:discussion}.

\section{Maximum Gravity Model}
\label{sec:MG-model}
\subsection{Predictions}
\begin{figure}
    \centering
    \includegraphics[width=0.495\textwidth]{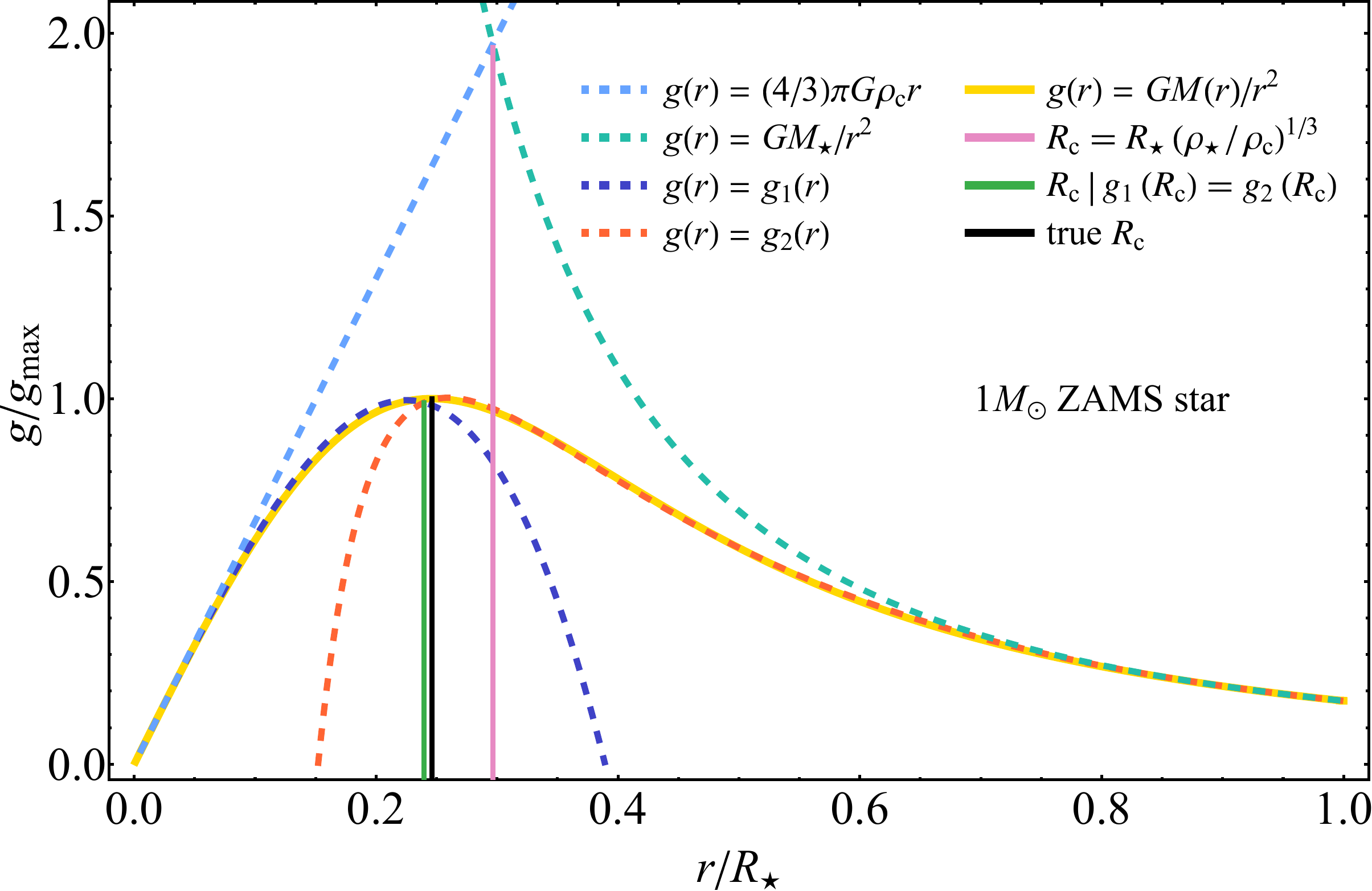}
    \caption{The self-gravitational field $g(r)$ (yellow) of a 1.0$\Msun$ ZAMS star, normalized by its maximum value $g_{\rm max}$. The solid black line shows the core radius $R_{\rm c}$ at which $g(r)$ is maximized. Also shown are the leading order approximations for $g$ close to the center (blue dashed) and the surface (teal dashed) of the star, and the next to leading order approximations ($g_1$ and $g_2$ shown by the purple and orange dashed curves). The pink and green solid lines show the intersection of the leading order and next to leading order approximations, respectively, which yield estimates for the core radius $R_{\rm c}$.} 
    \label{fig:maximums}
\end{figure}
The MG model postulates that a star is completely destroyed when the tidal field of an SMBH exceeds the maximum value of the self-gravitational field, $g_{\rm max}$, of the star. The radius within the star at which its gravitational field is maximized is denoted as the ``core radius,'' $R_{\rm c}$. The location of the core radius depends on the stellar profile, and in general can be straightforwardly and numerically evaluated. However, directly solving for the core radius for a given stellar structure does not lend obvious insight into its variation with the bulk properties of a star, such as $M_\star$ and $R_\star$. Alternatively, an estimate for the core radius $R_{\rm c}$ can be obtained by using the lowest-order expansions for the gravitational field of the star near its center ($g(r) = (4/3)\pi G \rho_{\rm c} r$) and surface ($g(r) = G M_\star/r^2$).~\cite{coughlin22a} demonstrated that setting the two equal to one another yields a parameterization of the core radius in terms of the bulk properties of the star, namely $\rho_{\rm c}, M_\star$ and $R_\star$, namely
\begin{equation}
    R_{\rm c} = R_\star\left( \frac{\rho_\star}{\rho_{\rm c}}\right)^{1/3}, \label{eq:Rc_LO}
\end{equation}
where $\rho_\star = 3M_\star/4\pi R_\star^3$ is the average stellar density. Figure~\ref{fig:maximums} shows the lowest-order approximations for $g(r)$ and the approximate core radius $R_{\rm c}$ (given by Equation~\ref{eq:Rc_LO}) at which the two curves intersect, for a $1M_\odot$ ZAMS star, in relation to their exact values as obtained from the {\sc mesa}~\citep{mesa} profile. As seen in the figure, the approximate curves intersect at a larger radius than the true core radius, and yield a $g_{\rm max}$ that is $\sim$ twice its true value.
\begin{figure}
    \centering
    \includegraphics[width=0.495\textwidth]{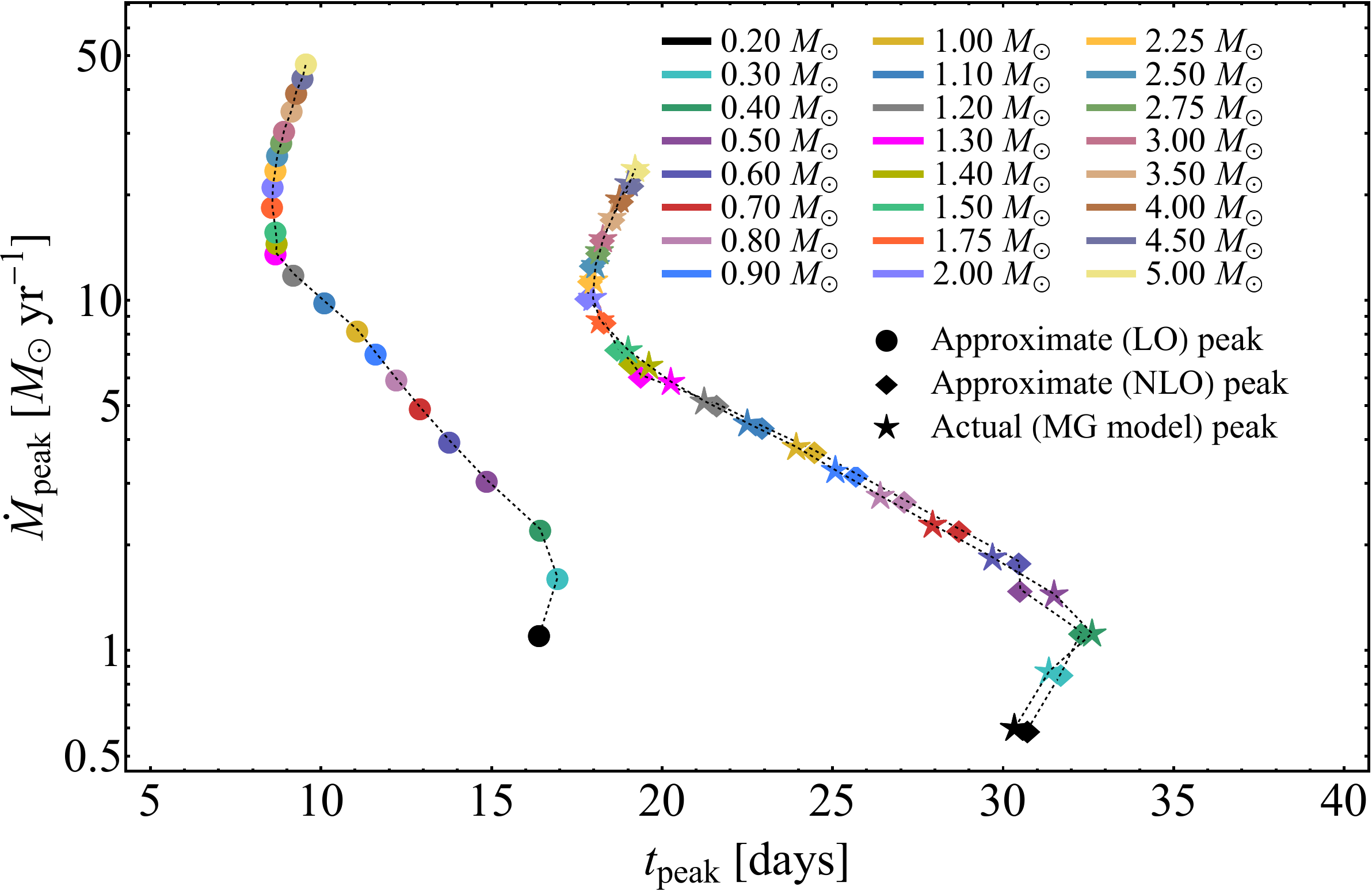}
    \caption{The peak fallback times $t_{\rm peak}$ and rates $\dot{M}_{\rm peak}$ for $0.2-5M_\odot$ ZAMS stars disrupted by a $10^6M_\odot$ SMBH. The 5-pointed stars represent the exact MG prediction, obtained by using the true location of the core radius $R_{\rm c}$. The circles and diamonds represent the predicted peaks using the leading order (LO) and next to leading order (NLO) approximations for the self-gravitational field, and the corresponding values of $R_{\rm c}$. This demonstrates that, while the leading order expansion provides a significant improvement on the frozen-in approximation, the next terms in the expansion of the gravitational field are required to accurately predict the peak fallback rate properties.}
    \label{fig:peak-timescales}
\end{figure}

\begin{figure*}
    \centering
    \includegraphics[width=0.49\textwidth]{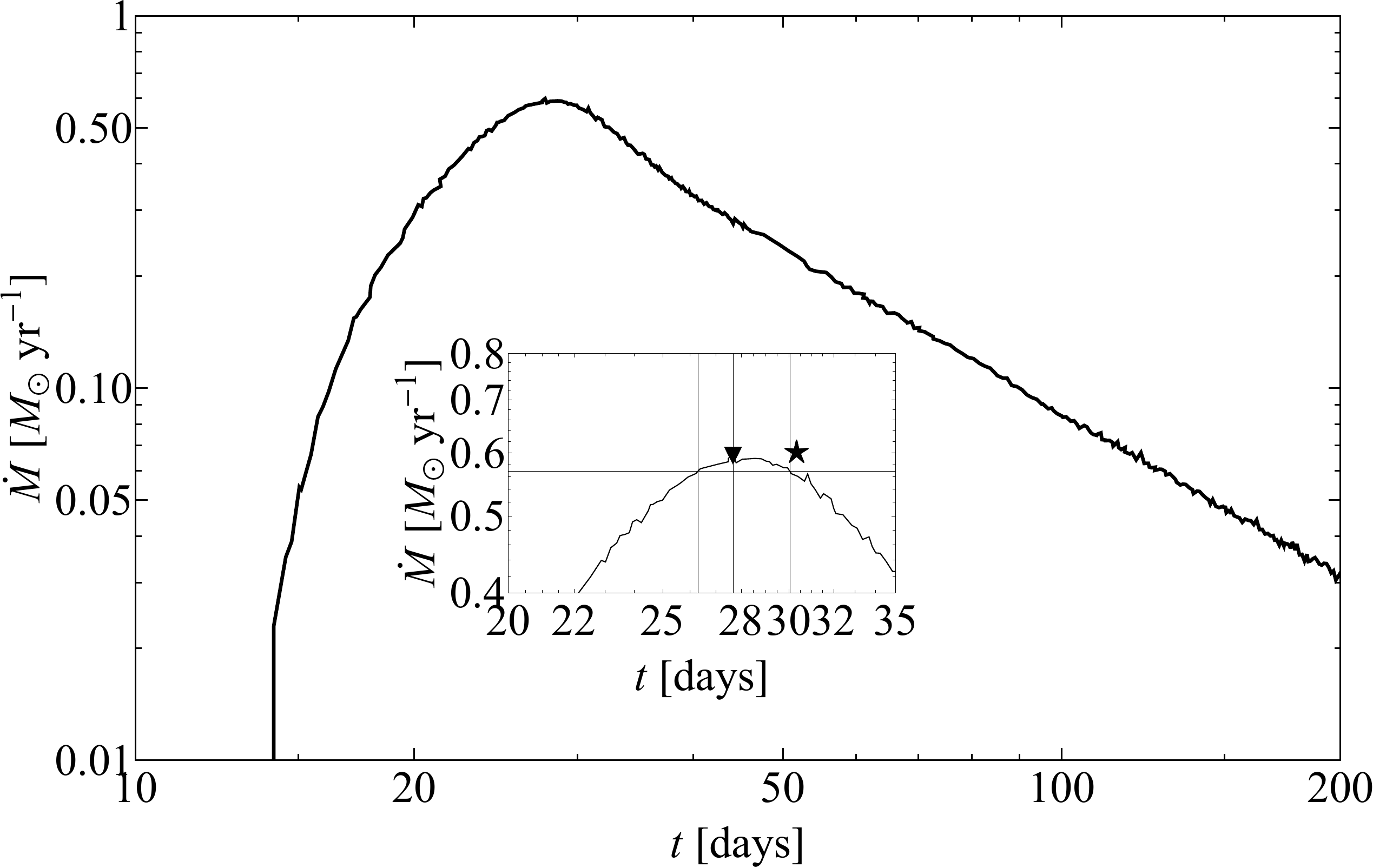}
    \includegraphics[width=0.49\textwidth]{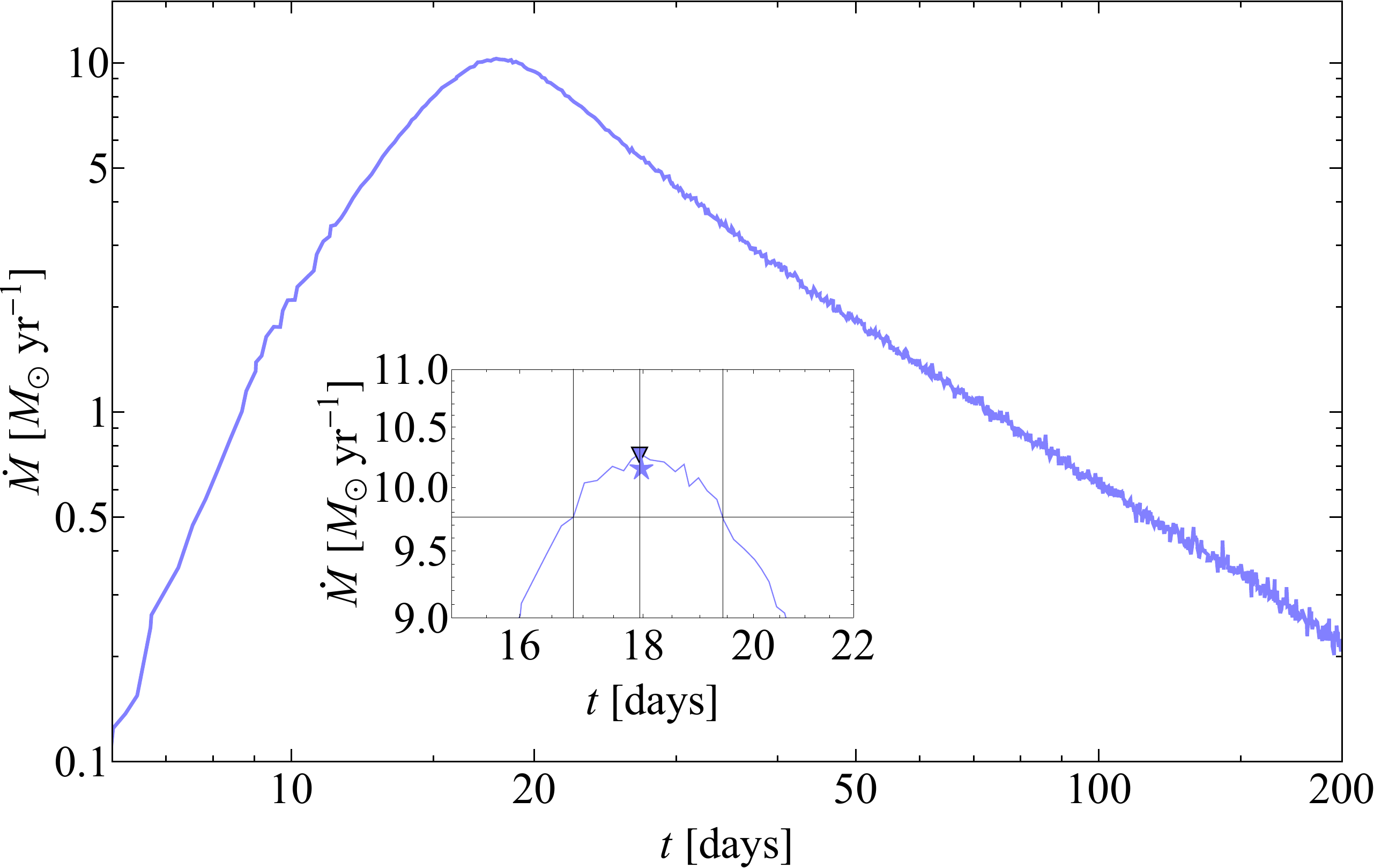}
    \caption{The fallback curves following the disruption of a 0.2$\Msun$ ZAMS star (left) and a 2.0$\Msun$ ZAMS star (right) by a $10^6M_{\odot}$ SMBH. The insets zoom in on the peak, where the inverted triangle shows the numerical peak value and the star marks the MG model prediction. The horizontal line shows 95$\%$ of the peak value, and the vertical lines show the time at which the fallback rate reaches 95$\%$ of the peak value, represented in Figure \ref{fig:zamsplot} as the error bars.}
    \label{fig:zamsfallbackrates}
\end{figure*}

We can obtain a more accurate estimate of $R_{\rm c}$ by incorporating additional details of the stellar structure. To do so, we parameterize the approximate expressions for the self-gravitational field within the star in terms of higher order moments of its mass distribution. The mass contained within a small volume of radius $R$ near the center of the star can be estimated by using the Taylor expansion of the density about $R=0$:
\begin{equation}
    M_1(R)=4\pi\left(\rho_{\rm{c}}\frac{R^3}{3}-\frac{1}{2}\rho_{\rm{c}}'' \frac{R^5}{5} \right),
\end{equation}
where $\rho_{\rm{c}}''$ is the second derivative of the density evaluated at the center of the star, such that the self-gravitational field is
\begin{equation}
    g_1(R)=4\pi G\left(\rho_{\rm{c}}\frac{R}{3}-\frac{1}{2}\rho_{\rm{c}}'' \frac{R^3}{5} \right).
\end{equation}
Conversely, near the stellar surface the mass enclosed is
\begin{equation}
    M_2(R)= M_\star \left[ 1+A\left(1-\frac{R}{R_\star} \right)^n \right],
\end{equation}
where $A$ and $n$ are constants that can be determined from the {\sc mesa}-generated mass profile (for the $1M_\odot$ ZAMS star the best fit values are $A=-1.8$ and $n=3.6$), such that 
\begin{equation}
   g_2(R)= \frac{GM_\star}{R^2} \left[ 1+A\left(1-\frac{R}{R_\star} \right)^n \right]
\end{equation}
is the corresponding self-gravitational field. The purple- and orange-dashed curves in Figure~\ref{fig:maximums} show the functions $g_1$ and $g_2$ for the $1M_\odot$ ZAMS star, and the green solid line marks the intersection of the two curves. As seen in the figure, the core radius $R_{\rm c}$ obtained from the intersection of the $g_1$ and $g_2$ curves very closely approximates the true core radius (shown by the solid black line). 

The approximate expressions for the gravitational field can be used to determine the critical tidal disruption radius radius $r_{\rm{t,c}}$ (i.e., the distance of closest approach at which the star would be completely destroyed by tides) by equating the tidal field of the black hole at the location of the core to the star's self-gravitational field at that location. We can then calculate the peak timescale $t_{\rm peak}$ and the peak fallback rate $\dot{M}_{\rm peak}$ using Equations 11 and 12 in~\cite{coughlin22a}. Setting the tidal field of the black hole equal to $(4/3) \pi G \rho_{\rm c} R_{\rm c}$ (the lowest order approximate expression for the gravitational field at the core) yields, for the critical tidal radius,
\begin{equation}
    r_{\rm t,c} = r_{\rm t} \left( \frac{\rho_{\rm c}}{ 4 \rho_{\star}}\right)^{-1/3},
\end{equation}
where $r_{\rm t}$ is the canonical tidal radius. The peak timescale (using Equation 11 of~\citealt{coughlin22a}) then evaluates to
\begin{equation}
    t_{\rm peak} = \left( \frac{\rho_{\rm c}}{4 \rho_{\star}} \right)^{-1/2}  \left( \frac{r_{\rm t}^2}{2 R_{\star}}\right)^{3/2} \frac{2 \pi }{\sqrt{G M_{\bullet}}}. \label{eq:tpeak-approx}
\end{equation}
In addition to the fallback timescale, $t_{\rm fb} = \left( r_{\rm t}^2/2R_\star\right)^{3/2} \times 2\pi/\sqrt{G M_\bullet}$~\citep{lacy82}, the expression for $t_{\rm peak}$ depends on the ratio $\rho_{\rm c}/\rho_\star$ of the central density to the average stellar density. The corresponding peak fallback rates can be expressed as $\dot{M}_{\rm peak} = M_\star/4t_{\rm peak}$ (using Equation 12 of~\citealt{coughlin22a}). 

Figure~\ref{fig:peak-timescales} shows the approximate peak timescales (depicted with circles) obtained using Equation~\eqref{eq:tpeak-approx}, and peak fallback rates for $0.2-5M_\odot$ ZAMS stars, alongside the $(t_{\rm peak},\dot{M}_{\rm peak})$ values obtained using the exact core radius $R_{\rm c}$ (depicted with 5-pointed stars), for tidal disruption by a $10^6 M_\odot$ SMBH. As seen in the figure, the approximate expression for $t_{\rm peak}$ underpredicts its true value by a factor of $\sim2-3$ for the ZAMS stars, and overpredicts $\dot{M}_{\rm peak}$.~\cite{bandopadhyay24} showed that the peak timescale is roughly independent of stellar mass and age, and for a $10^6 M_\odot$ SMBH, ranges between $\sim 15-30$ days. We can thus infer that the approximate peak timescale given by Equation~\eqref{eq:tpeak-approx}, which scales as $\propto \left( \rho_{\rm c}/4 \rho_{\star} \right)^{-1/2}$, becomes increasingly discrepant for highly evolved stars, for which the ratio of the central density to the average density increases. 

Figure~\ref{fig:peak-timescales} also shows the approximate $t_{\rm peak}$ and $\dot{M}_{\rm peak}$ obtained using the next-to leading order approximations for the self-gravitational field for the same range of stars. The agreement between the higher-order-approximated $t_{\rm peak}$ and $\dot{M}_{\rm peak}$ and the exact values (i.e., using the exact values of $R_{\rm c}$ and $r_{\rm t,c}$) indicates that higher-order moments of the mass distribution of a star are necessary to accurately parameterize the peak fallback properties. Conversely, accurately predicting the peak fallback rate properties cannot be achieved with only $\rho_{\rm c}, M_\star$ and $R_\star$. While this demonstrates the need for additional parameters to accurately model the peak of the fallback rate, it is more straightforward to simply use the exact values of $R_{\rm c}$ and $r_{\rm t,c}$; in the next subsection we compare these values to those obtained from numerical simulations of TDEs.

\begin{figure*}
    \centering
    \includegraphics[width=1.0\textwidth]{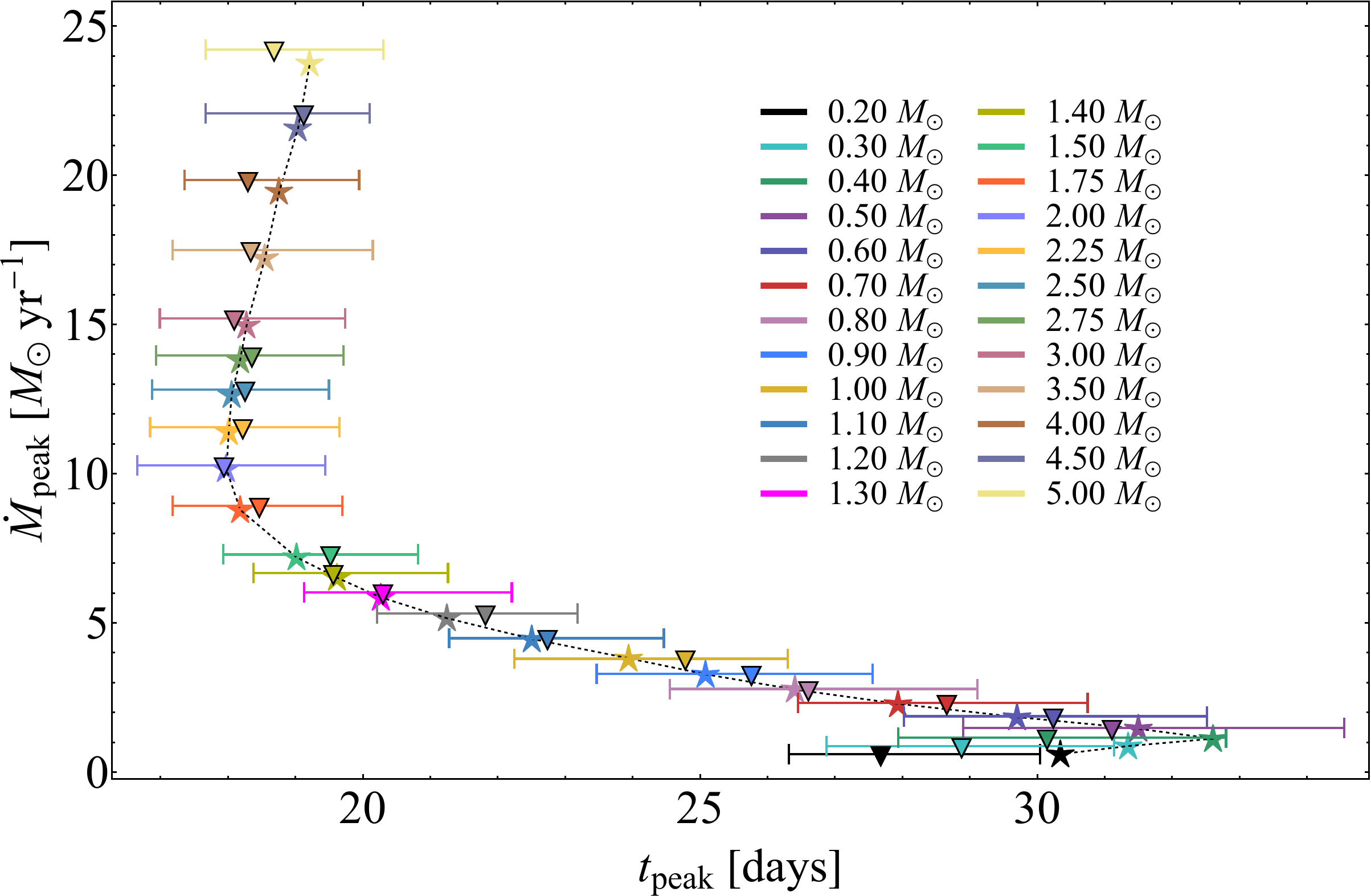}
    \caption{The predictions of the MG model for the disruption of ZAMS stars at varying masses (stars) compared to the numerical value of the peak fallback rate (triangles). The horizontal lines provide the amount of time for which the fallback rate is above $95\%$ of its peak value; see the insets in Figures \ref{fig:zamsfallbackrates} and \ref{fig:mams+tamsfallbackrates} for examples.}
    \label{fig:zamsplot}
\end{figure*} 

\begin{figure*}
    \centering
    \includegraphics[width=0.49\textwidth]{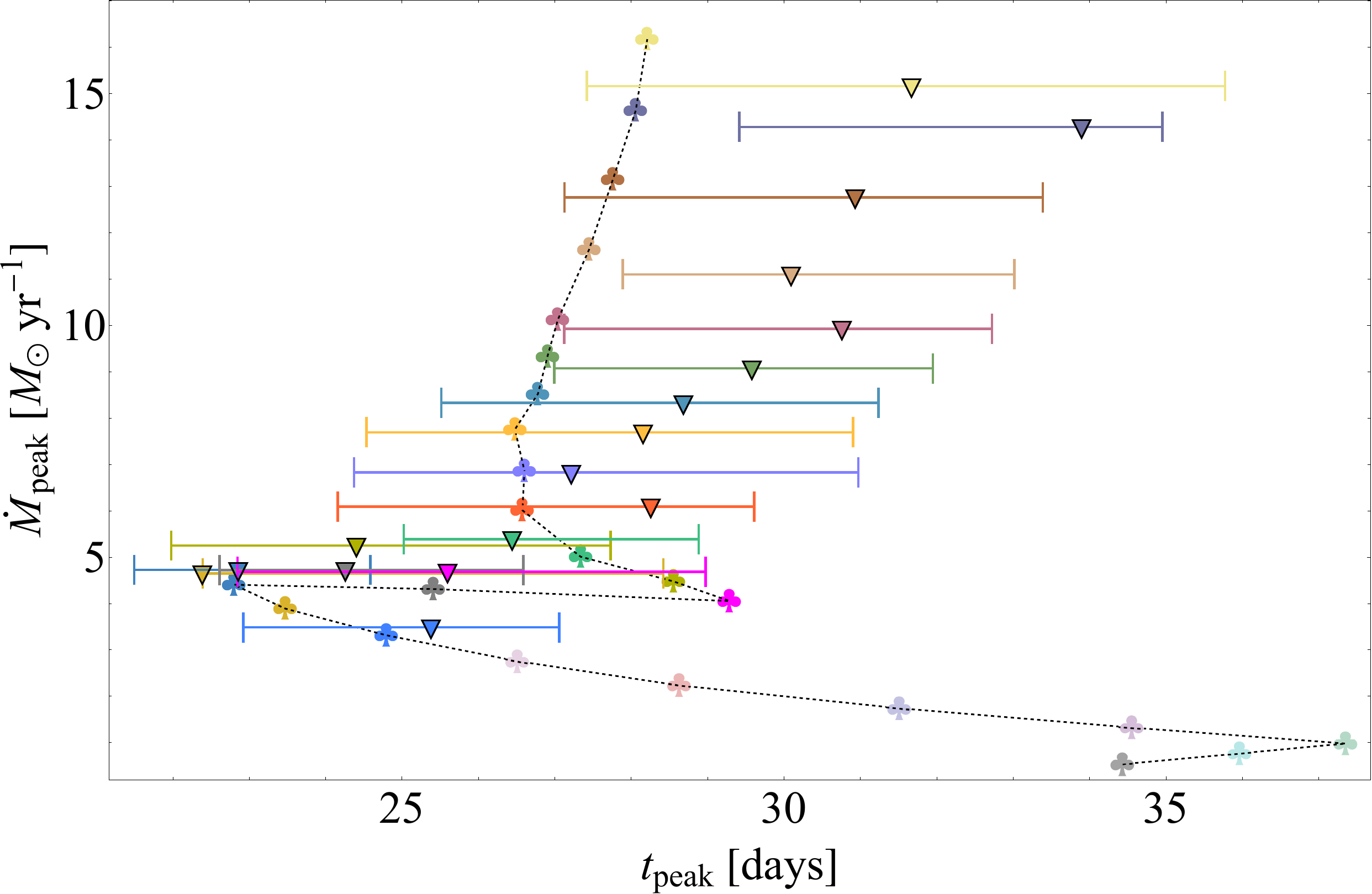}
    \includegraphics[width=0.49\textwidth]{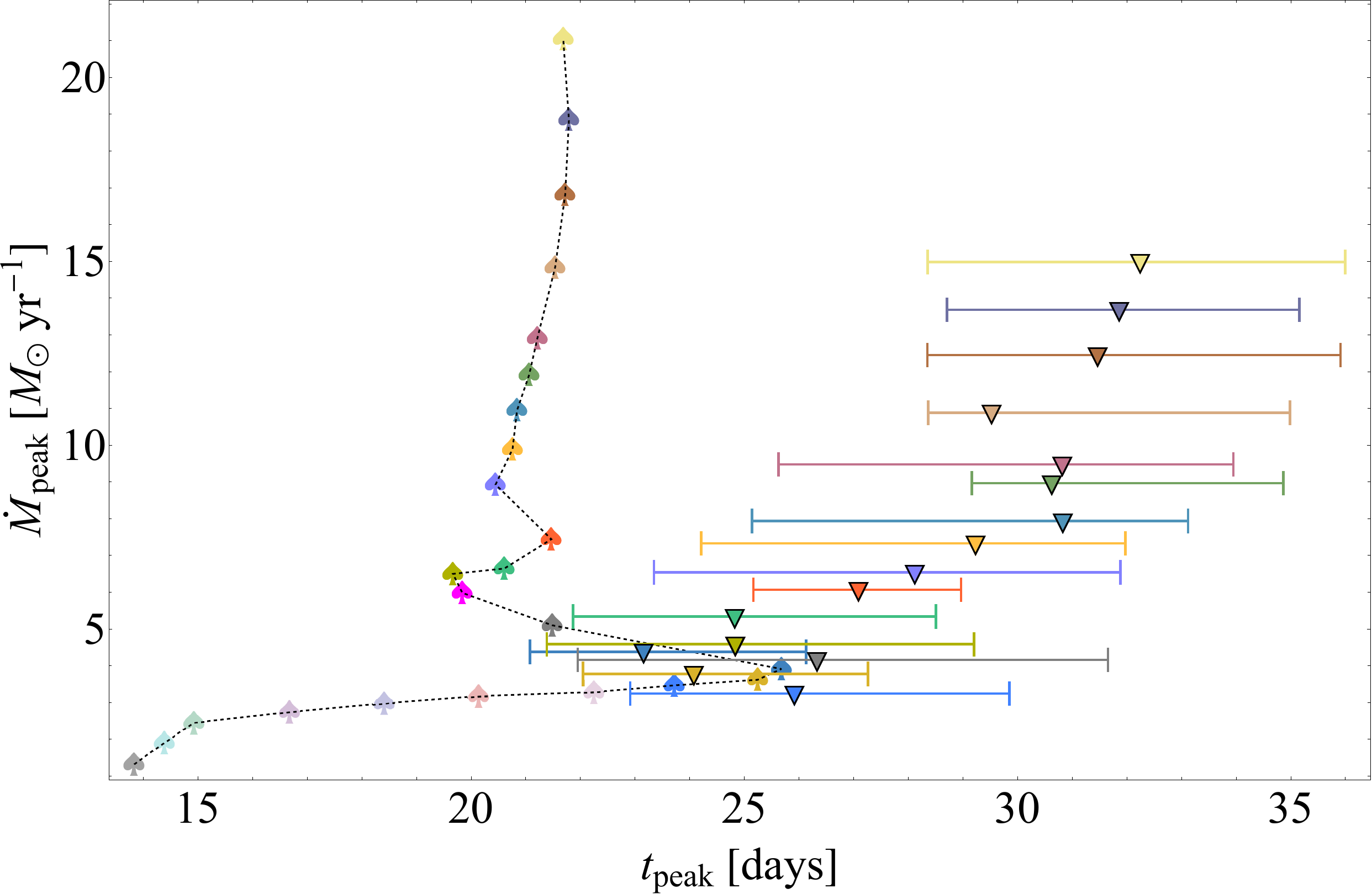}
    \caption{Same as Figure \ref{fig:zamsplot} but for the disruption of MAMS (left) and TAMS (right) stars.}
    \label{fig:mams+tamsplot}
\end{figure*}

\subsection{Numerical Simulations}

We test the MG model by running multiple sets of numerical simulations of TDEs. We evolve stars with masses ranging from 0.2--5.0$\Msun$ through the main sequence using \textsc{mesa} \citep{mesa}. The stellar profiles are then imported into the smoothed particle hydrodynamics (SPH) code \textsc{phantom} \citep{phantom}, where they are relaxed for $\sim$ 5 sound crossing times using the routine implemented in \cite{golightly19,nixon21}. We use a resolution of $10^6$ particles to model the stars in {\sc phantom}. We simulate the disruption of stars by a $10^6M_\odot$ SMBH modeled as a Newtonian point mass potential, with the relaxed \textsc{mesa} star initially placed at a distance of 5$r_{\rm{t}}$ on a parabolic Keplerian orbit with pericenter $r_{\rm t, c} = r_{\rm t}/\beta_{\rm c}$, with $\beta_{\rm c}$ determined from the MG model. We run three sets of simulations of 24 stars in the mass range 0.2--5.0$\Msun$. The three sets are at different ages: zero-age main sequence (ZAMS), terminal-age main sequence (TAMS), and ``middle-age main sequence" (MAMS), defined as the time at which the central hydrogen mass fraction falls below 0.2. We do not simulate MAMS and TAMS stars with masses below 0.9$\Msun$, as stars of these masses would not reach MAMS or TAMS in the age of the universe.

The fallback rates are calculated directly in \textsc{phantom} by counting particles as they pass into the accretion radius of the SMBH (we find that the bin size used to calculate the fallback rate does not significantly impact the peak magnitude or peak timescale). Approximately one day after the star passes pericenter, the accretion radius is increased to 120 gravitational radii \citep{coughlin15,golightly19}. If a core reforms within that same time period, the core is replaced by a sink particle, which is a good approximation for the core (for more detail, see \citealt{miles20,nixon21,bandopadhyay24b}).

The directly calculated fallback rates for a 0.2$\Msun$ (left) and a 2.0$\Msun$ star at ZAMS (right) are shown in Figure \ref{fig:zamsfallbackrates}. The inset shows a zoomed-in view of the peak, where the numerical peak is marked with an inverted triangle. The MG prediction is marked with a star. The horizontal line delimits where the fallback rate is 95$\%$ of it peak magnitude, and the outer vertical lines mark the times at which the fallback rate is $95\%$ of the peak value. The central vertical line marks the time at which the peak occurs. The MG prediction for the $0.2\Msun$ ZAMS star falls slightly outside the 95$\%$ interval because the peak is much shallower and slight numerical noise has influenced the value chosen as the peak. 
\begin{figure*}
    \centering
    \includegraphics[width=0.49\textwidth]{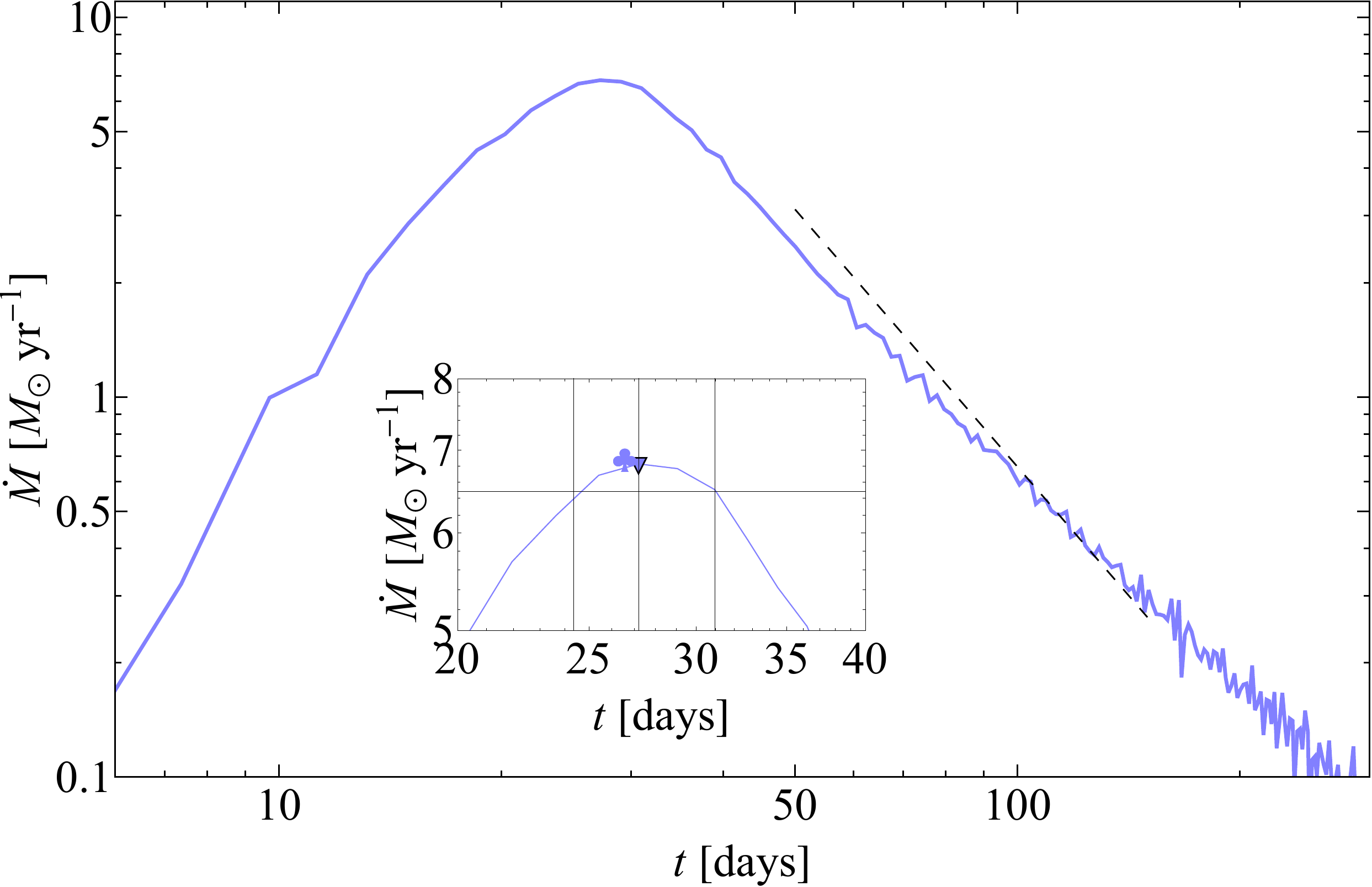}
    \includegraphics[width=0.49\textwidth]{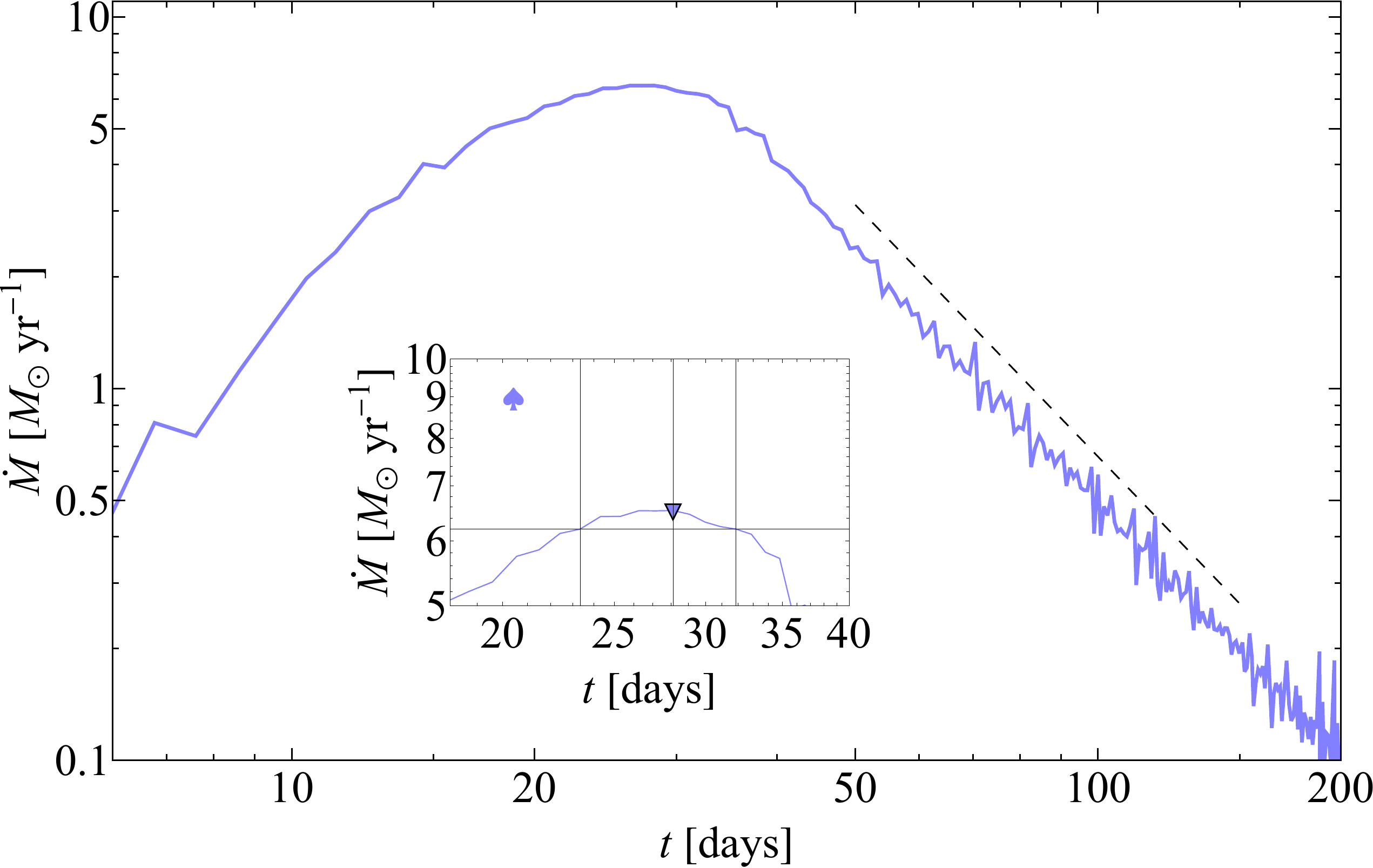}
    \caption{Same as Figure \ref{fig:zamsfallbackrates}, but for the disruption of a 2.0$\Msun$ star at MAMS (left) and TAMS (right). The black-dashed line shows a scaling of $t^{-9/4}$, demonstrating that these are partial disruptions.} 
    \label{fig:mams+tamsfallbackrates}
\end{figure*}
Figure \ref{fig:zamsplot} shows the comparison between the predicted peak fallback rate of the MG model (shown as stars) and the numerical peak fallback rate from the set of simulations using ZAMS stars (shown as inverted triangles), where the color represents the mass of the star. The error bars mark the duration of time for which the fallback rate is above $95\%$ of its peak value, as shown by the insets of Figure \ref{fig:zamsfallbackrates}. The numerical values thus agree extremely well with the MG prediction over the entire mass range.

Figure \ref{fig:mams+tamsplot} shows the results from simulations of MAMS (left) and TAMS (right) stars compared with the MG model predictions. The error bars, as in Figure \ref{fig:zamsplot}, mark the duration for which the fallback rate is above $95\%$ of $\dot{M}_{\rm{peak}}$. It is clear that the model is less accurate for evolved stars, and the discrepancy between the prediction and the numerically obtained results increases as mass and age increase. Nonetheless, the MG model is still considerably more accurate than the frozen-in approximation, which overestimates the peak timescale by months to years and underestimates the peak fallback rate by up to two orders of magnitude for the highest-mass stars \citep{golightly19,bandopadhyay24}.

There are two possible origins\footnote{In addition to the two physical origins we describe, a third possibility could be related to a lack of numerical convergence, but we directly demonstrate in Appendix~\ref{sec:convergence} that our results are strongly converged -- especially near the peak in the fallback rate.} for the discrepancy between the predictions of the MG model and the {\sc phantom} fallback rates from evolved stars. First, if the disruption is only partial, the presence of a core modifies the debris dynamics~\citep{guillochon13,coughlin19}, and a core reforms almost immediately after pericenter for the MAMS and TAMS stars. Figure \ref{fig:mams+tamsfallbackrates} shows the fallback rates of a $2.0\Msun$ star at MAMS (left) and TAMS (right), where the insets are the same as in Figure \ref{fig:zamsfallbackrates}, but the MG model predictions are shown by the clover and spade respectively. The fallback rates exhibit a $t^{-9/4}$ scaling (the black dashed line in the figure shows a $\propto t^{-9/4}$ power law) at late times, appropriate to the case of a partial disruption~\citep{coughlin19}. The MG model assumes that the star is completely destroyed, and is hence less accurate for the case of these partial disruptions. 

Alternatively, the core radius itself may differ from the model prediction, which could be due to the tidal compression of the star and the resultant increase in the central density and self-gravitational field.\footnote{In support of the fact that self-gravity is significantly increased in deep encounters, \citet{nixon22} found that a $5/3$ polytropic star is completely destroyed by a $10^6 M_{\odot}$ SMBH for $\beta = 8$, while a core reforms for $\beta = 16$.} In particular, the MG model assumes that the maximum self-gravitational field against which the tidal field competes is established by the original density profile of the star, but for deep encounters -- which are required to completely destroy high-mass and evolved stars -- the central density (and self-gravitational field) is augmented by the vertical component of the SMBH tidal field \citep{carter82}, thus making the core more capable of withstanding the tidal shear (and making the core radius smaller). A smaller $R_{\rm c}$ than predicted would yield a longer peak timescale (using Equation 11 of \citealt{coughlin22a}) and, correspondingly, a smaller $\dot{M}_{\rm peak}$. To test which of these possibilities -- the reformed core or a smaller $R_{\rm c}$ -- is more responsible for modifying the values from their MG predictions, we performed additional simulations at $\beta=\beta_{\rm c}+1$, and found that the MAMS stars are completely destroyed at $\beta_{\rm c}+1$ (with the exception of the $0.9M_\odot$ MAMS star, which is still partially disrupted). Despite not having a core, these disruptions still yield fallback rates with longer peak timescales and lower peak magnitudes compared to the model predictions. This finding suggests that the core radius is smaller than that predicted by the MG model and is primarily responsible for introducing the discrepancies present in Figure \ref{fig:mams+tamsplot}, but that the $\beta_{\rm c}$ predicted by the model is still accurate.

Because the tidal compression strengthens the self-gravitational field, it may be impossible to completely destroy high-mass TAMS stars with a Newtonian tidal field. However, this resistance to disruption may be unphysical, because relativistic effects (which strengthen the tidal field of the black hole) become more pronounced at such high $\beta$'s. In the next section we construct a simple relativistic generalization of the MG model to address the importance of such effects.

\section{Relativistic Generalization}
\label{sec:pw-potential}
\begin{figure*}
    \includegraphics[width=0.495\textwidth]{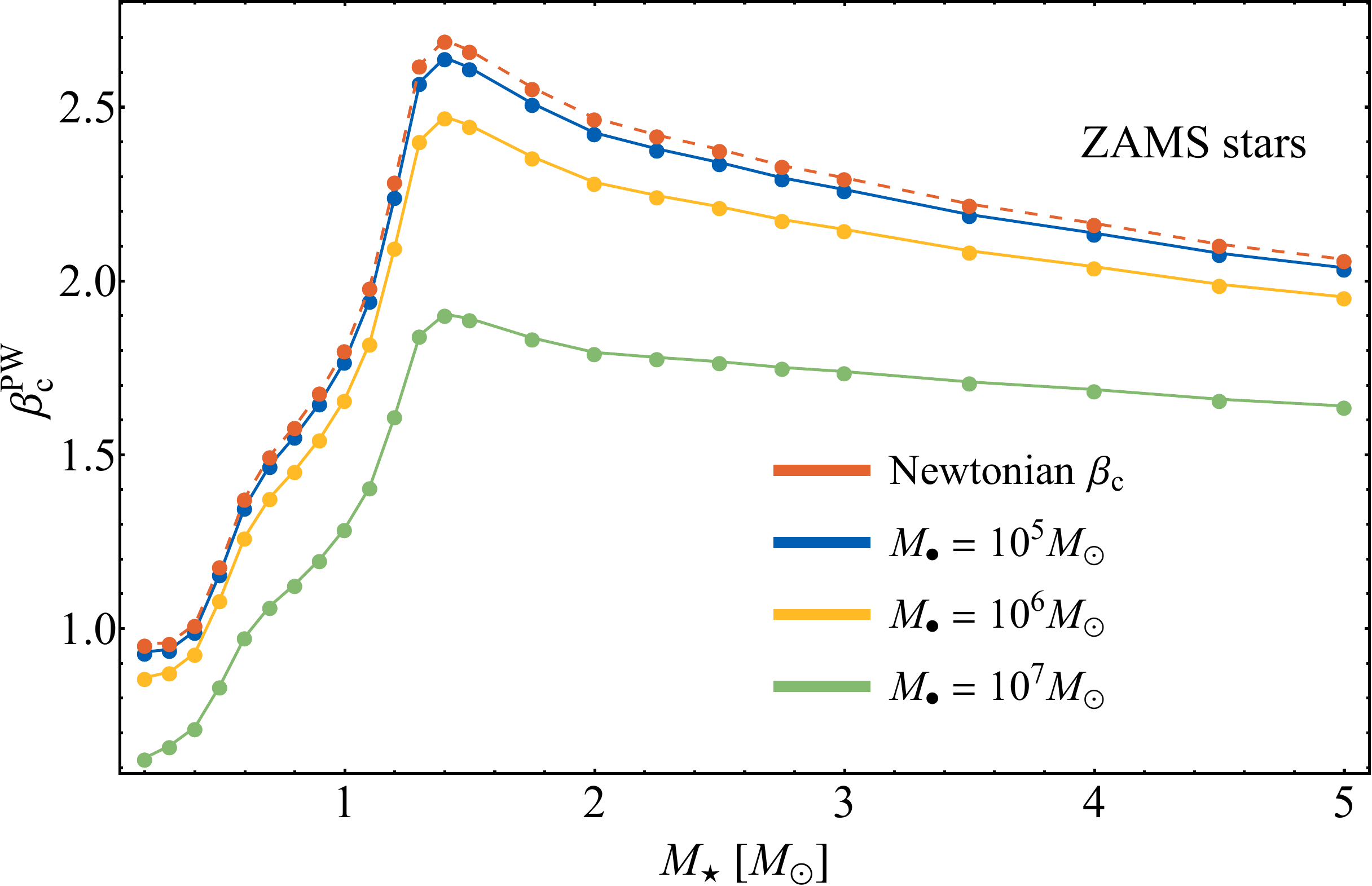} 
    \includegraphics[width=0.51\textwidth]{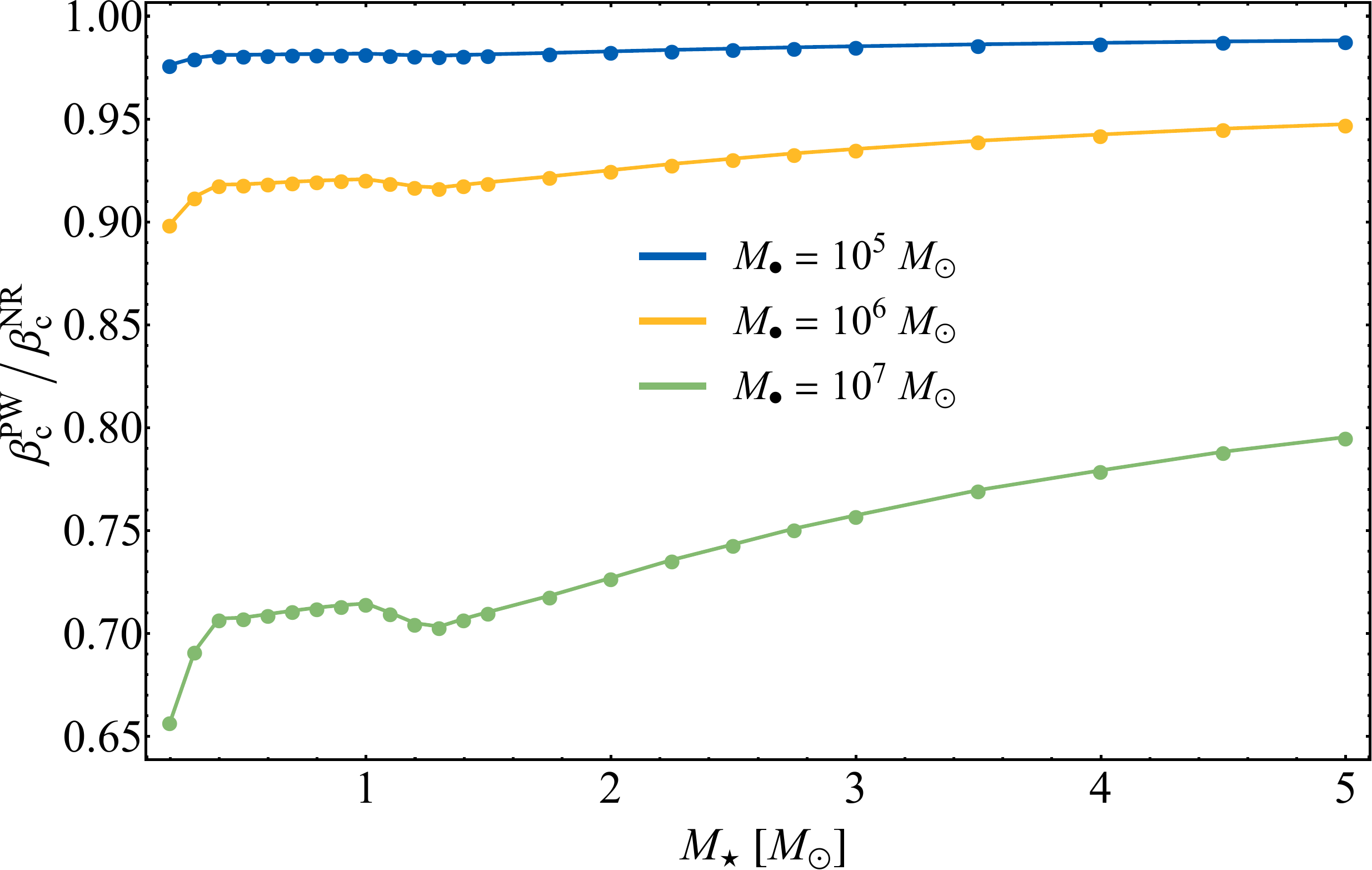} 
    \caption{Left: The critical impact parameter $\beta_{\rm c}^{\rm PW}$ obtained using the Packzy{\'n}ski-Wiita potential, as a function of stellar mass, for zero-age main sequence stars disrupted by SMBHs of masses $10^5, 10^6$ and $10^7M_{\odot}$ (represented by the the blue, yellow and green lines respectively). The orange dashed curve shows the $\beta_{\rm c}$ estimated using the Newtonian potential for the tidal field of the SMBH (identical to the ZAMS curve in the left panel of Figure 4 of~\citealt{bandopadhyay24}), which is independent of the SMBH mass}. Right: The ratio of $\beta_{\rm c}$ for the Packzy{\'n}ski-Wiita potential ($\beta_{\rm c}^{\rm PW}$) to the non-relativistic value ($\beta_{\rm c}^{\rm NR}$), for the same range of ZAMS stars and three different SMBH masses. 
    \label{fig:pw-betas}
\end{figure*}

\begin{figure*}
    \includegraphics[width=0.51\textwidth]{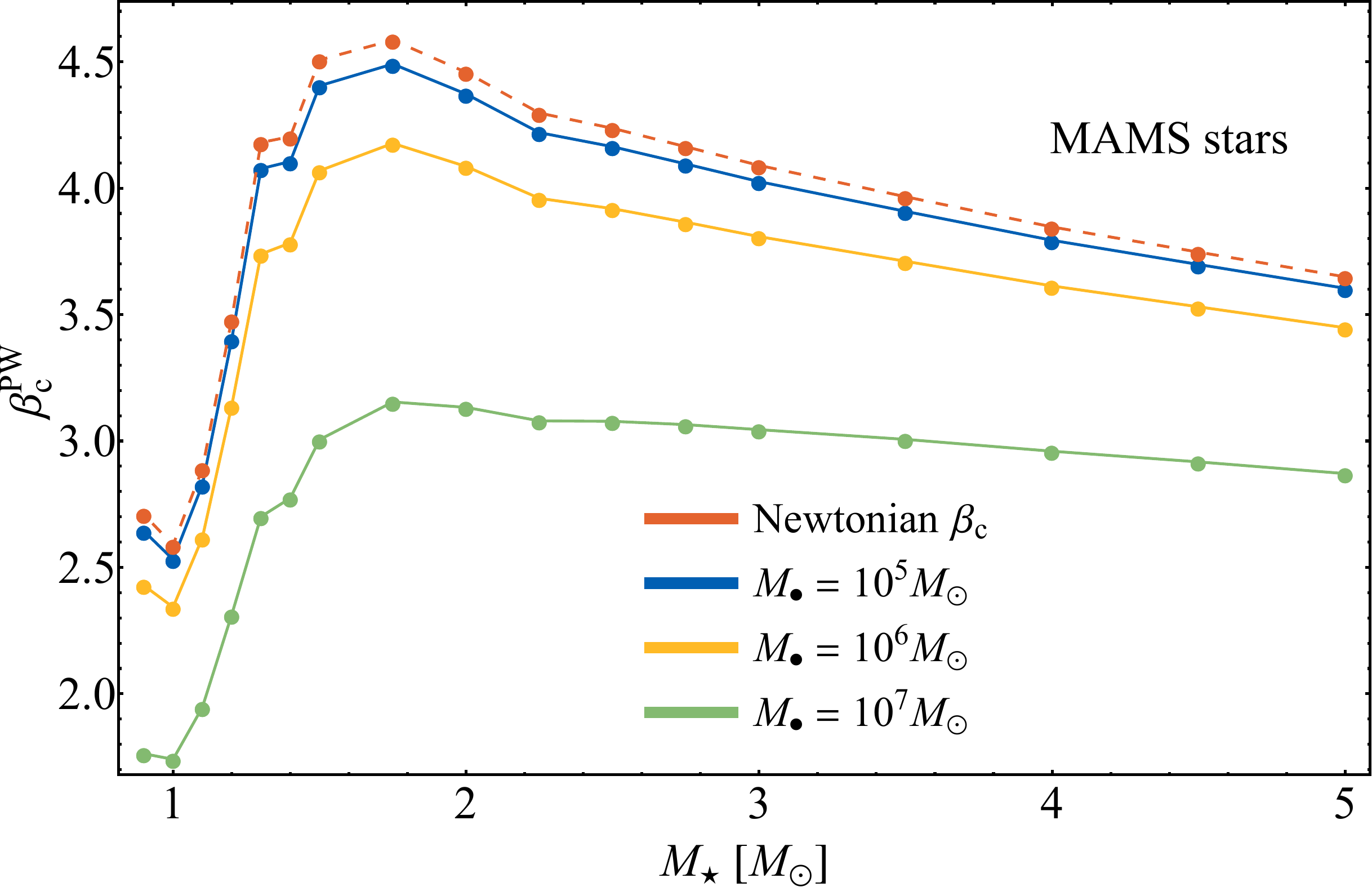} 
    \includegraphics[width=0.495\textwidth]{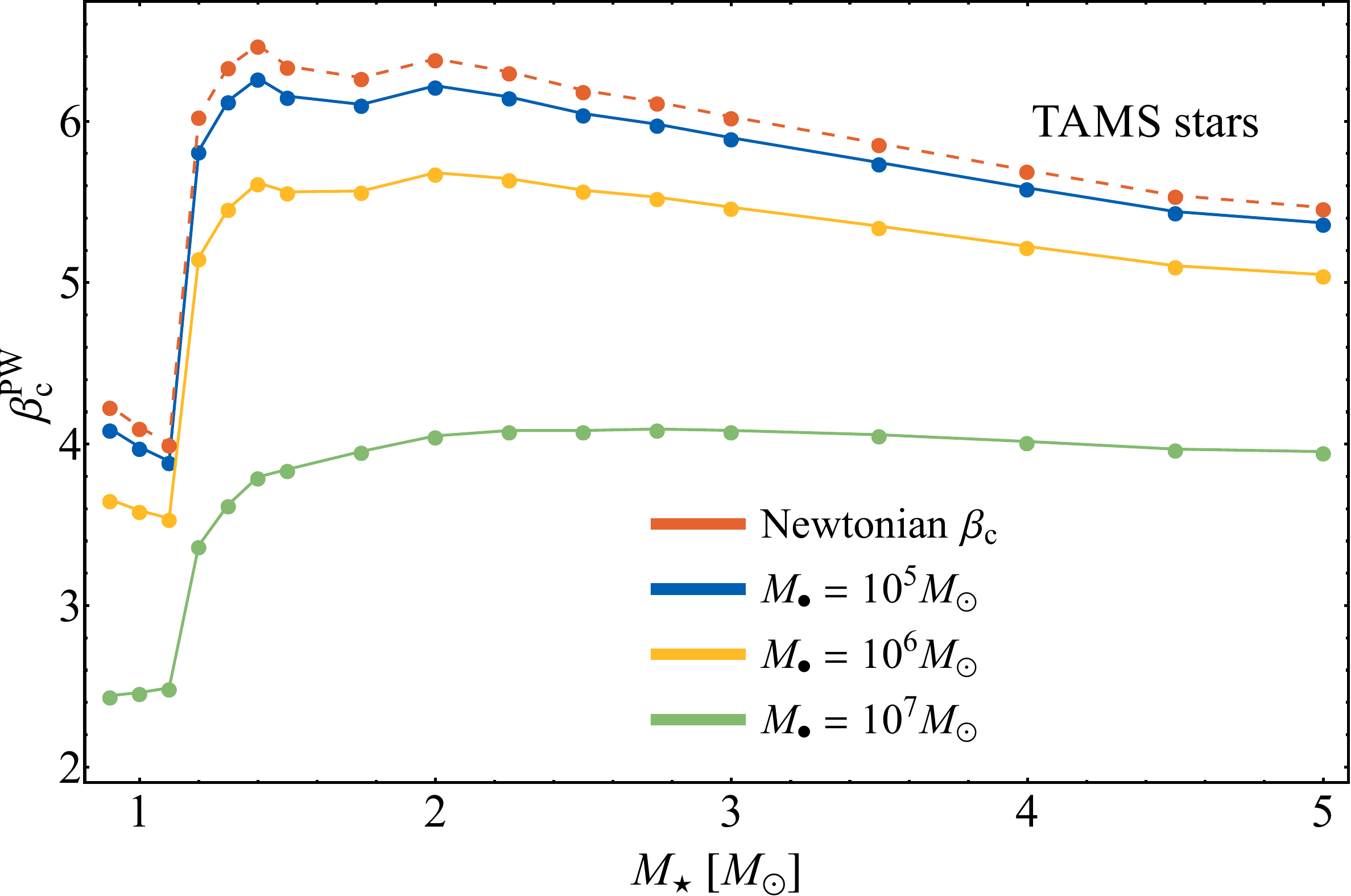} 
    \caption{Same as Figure \ref{fig:pw-betas}, but for the disruption of middle-age main sequence (left) and terminal-age main sequence (right) stars.}
    \label{fig:MAMS-and-TAMS-pw-betas}
\end{figure*}
The analytical model developed in~\cite{coughlin22a}  treats the SMBH as a Newtonian point mass potential. For a sun-like star ($M_\star = 1 M_\odot; R_\star = 1 R_\odot$) being disrupted by a $10^6 M_\odot$ SMBH, the ratio of the tidal radius $r_{\rm t}$ to the gravitational radius $r_{\rm g} = G M_\bullet/c^2$ of the SMBH is $\sim 50$. The lengthscales involved in this system thus justify the Newtonian approximation, and a general relativistic treatment would yield minor corrections to $t_{\rm peak}$ and $\dot{M}_{\rm peak}$, as demonstrated by~\citet{gafton19}. However, the ratio $r_{\rm t}/r_{\rm g}$ scales as $\propto M_{\bullet}^{-2/3}$, and is thus increasingly non-ignorable as the SMBH mass grows. 

Here we estimate the importance of relativistic effects in modifying the tidal field of an SMBH by using the Packzy{\'n}ski-Wiita potential~\citep{paczynsky80},
\begin{equation}
    \Phi_\bullet = - \frac{G M_\bullet}{r-2r_{\rm g}},
\end{equation}
where $M_\bullet$ is the SMBH mass, and $r_{\rm g} = GM_{\bullet}/c^2$ is its gravitational radius, in place of the Newtonian potential used in \citet{coughlin22a}. For the complete disruption of a star, the tidal field evaluated at its core radius $R_\mathrm{c}$ must exceed the self gravitational field of the core, $g(R_{\rm c})$. This yields, for the critical disruption radius $r_{\rm t,c}^{\rm PW}$,
\begin{equation}
    \frac{4 G M_\bullet R_\mathrm{c}}{(r_{\rm t,c}^{\rm PW}-2r_{\rm g})^3} = g(R_{\rm c}) \,\, \Rightarrow \,\, r_{\rm t,c}^{\rm PW} = 2 r_{\rm g} + \left(\frac{4 G M_{\bullet} R_{\rm c}}{g(R_{\rm c})} \right)^{1/3}, \label{eq:rtc}
\end{equation}
where the second term on the right-hand side of the second equality is the Newtonian $r_{\rm t, c}$. This shows that the tidal field of the SMBH is stronger than its Newtonian estimate, thus resulting in a larger complete disruption radius. Equation~\eqref{eq:rtc} can be used to define the relativistic analog of the critical impact parameter required for the complete disruption of a star:
\begin{equation}
    \beta_{\rm c}^{\rm PW} = r_{\rm t}/r_{\rm t,c}^{\rm PW}.
    \label{eq:pw-beta}
\end{equation}

The left panel of Figure~\ref{fig:pw-betas} shows the critical impact parameter required for the complete disruption of $0.2-5 M_\odot$ ZAMS stars by SMBHs ranging from $10^5-10^7M_{\odot}$, alongside the Newtonian estimate for $\beta_{\rm c}$ (which is independent of SMBH mass), while the right panel gives the ratio $\beta_{\rm c}^{\rm PW}/\beta_{\rm c}^{\rm NR}$. Since the relativistic tidal field is stronger for a more massive SMBH, the minimum distance within which a star must approach to be destroyed by tides is larger, thus yielding a smaller $\beta_{\rm c}^{\rm PW}$ for higher-mass SMBHs. For all three SMBH masses shown in the figure, $\beta_{\rm c}$ reaches a maximum at $M_\star = 1.4 M_\odot$. This peak is due to the fact that $\beta_{\rm c}$ is strongly correlated with the ``compactness'' of a star, defined as the ratio of the core radius $R_{\rm c}$ to the stellar radius $R_\star$. Specifically, since the canonical tidal radius $r_{\rm t} = R_\star \times (M_\bullet/M_\star)^{1/3}$, while the critical disruption radius $r_{\mathrm{t,c}} = R_{\rm c} \times (4 M_\bullet /M_{\rm c})^{1/3}$ (where $M_{\rm c}$ is the core mass), $\beta_{\rm c} \equiv r_{\rm t}/r_{\mathrm{t,c}} \propto (R_{\star}/R_{\rm c})\times (M_{\star}/M_{\rm c})^{-1/3}$. The core mass fraction $M_{\rm c}/M_\star$ is $\sim 1$ if one approximates the density of the core as equal to $\rho_{\rm c}$ out to $R_{\rm c}$ \citep{coughlin22a}, and thus has a relatively minor influence on the variation in $\beta_{\rm c}$ compared to the compactness parameter. The fact that $\beta_{\rm c}$ has a relative maximum at $M_\star = 1.4 M_\odot$ thus arises primarily from the fact that this star has the smallest value of $R_{\rm c}/R_\star$ among all the ZAMS stars. The left and right panels of Figure~\ref{fig:MAMS-and-TAMS-pw-betas} are analogous to the left panel of Figure \ref{fig:pw-betas} but for MAMS and TAMS stars.

\begin{figure}
    \includegraphics[width=0.51\textwidth]{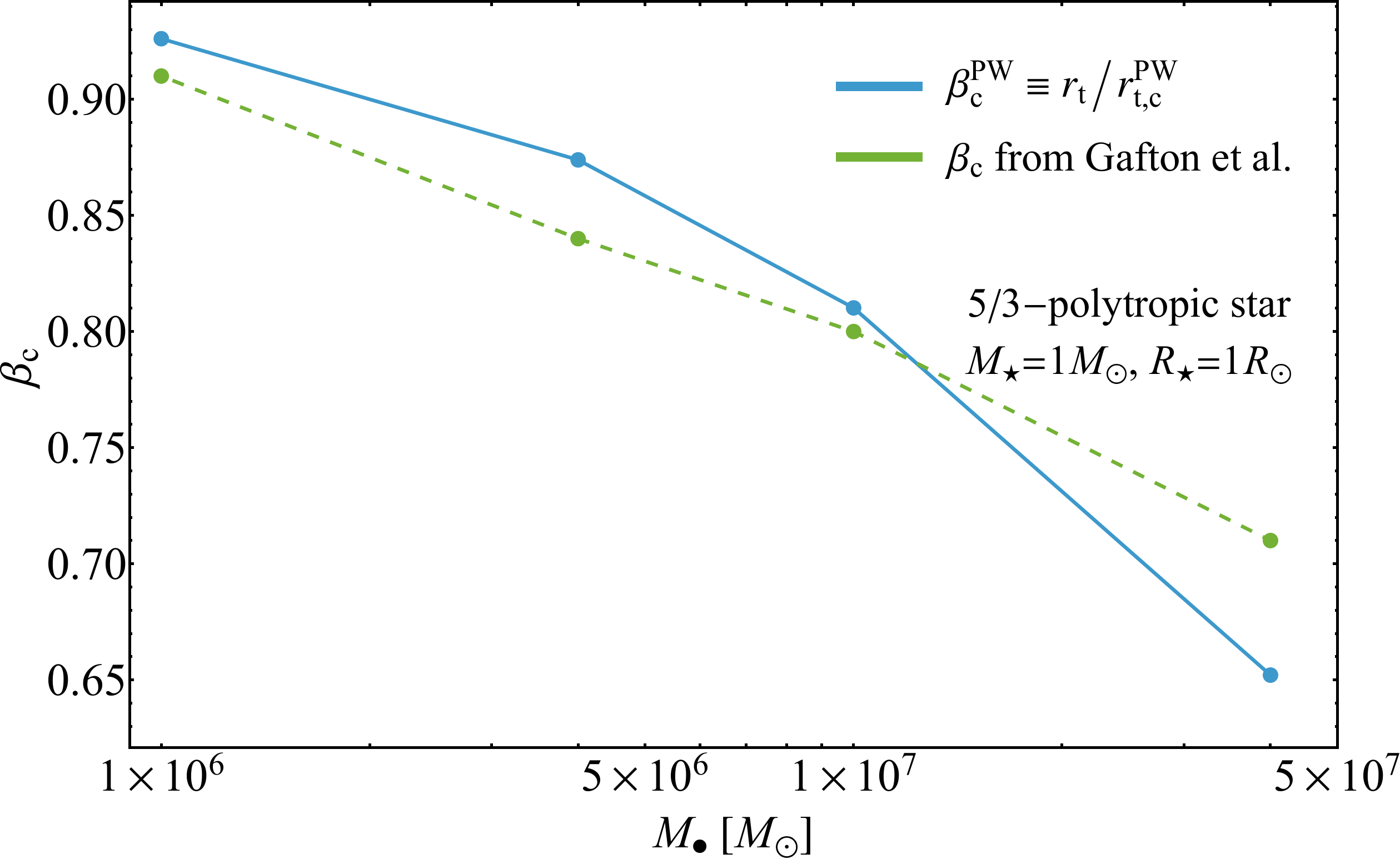} 
    \caption{The critical impact parameter predicted using the Packzy{\'n}ski-Wiita potential, required for the complete disruption of a sun-like star modeled as a $5/3-$polytrope (blue curve), as a function of SMBH mass, $M_\bullet$. Also shown are the results of numerical simulations performed in \cite{gafton15} (green curve) using the generalized Newtonian potential from~\cite{tejeda13} to account for relativistic effects. The points on the green curve correspond to the value of $\beta$ at which the core mass fraction goes to zero in Figure 3 of \cite{gafton15}. }
    \label{fig:MG-gafton-compare-betas}
\end{figure}

The estimate for the critical impact parameter $\beta_{\rm c}^{\rm PW}$ at which a star is completely destroyed is in good agreement with numerical simulations of TDEs that account for general relativistic effects (e.g.,~\citealt{gafton15,gafton19,ryu20c,jankovic23}). Specifically, for the ZAMS stars shown in Figure~\ref{fig:pw-betas}, the estimates for $\beta_{\rm c}^{\rm PW}$ are approximately equal to the values of $\beta$ in Figure 2 of \cite{jankovic23} at which the star is completely destroyed. Figure~\ref{fig:MG-gafton-compare-betas} shows a comparison of the critical impact parameter (using Eq~\ref{eq:pw-beta}) required for the complete disruption of a sun-like star modeled as a $5/3-$polytrope (solid blue curve), as a function of the SMBH mass, and the corresponding results from Figure 3 of~\cite{gafton15} (green dashed curve), who simulated the disruption of a $5/3-$polytropic star for a range of impact parameters and black hole masses. They used the generalized Newtonian potential derived in~\cite{tejeda13} to model the relativistic effects for a Schwarzschild black hole. As seen in the figure, the prediction obtained using the Packzy{\'n}ski-Wiita potential is in good agreement with the numerical results for black hole masses $M_\bullet \sim 10^6-10^7 M_\odot$. There is a larger discrepancy between the $\beta_{\rm c}$ predicted using this generalization for the MG model and the~\citeauthor{gafton15} result for $4\times 10^{7}M_{\odot}$, but for this SMBH mass the pericenter distance is $\sim 6 r_{\rm g}$ and therefore highly relativistic, where the Paczy\'nski-Wiita potential does not provide an accurate enough description of relativistic gravity. 

\section{Discussion and Conclusions}
\label{sec:discussion}
The TDE fallback rate -- the rate at which tidally stripped material returns to the disrupting black hole -- 
has been the subject of extensive analytical and numerical modeling. Here we performed hydrodynamical simulations to test the predictions of the maximum-gravity (MG) model, which was proposed in \citet{coughlin22a} and uses the core radius $R_{\rm{c}}$, being the location in the tidally destroyed star where its self-gravitational field is maximized, 
to predict the peak properties of the fallback rate. 
The predicted peak timescale $t_{\rm peak}$ and peak fallback rate $\dot{M}_{\rm peak}$ agree exceptionally well with the results of numerical simulations for the complete disruption of ZAMS stars (as shown in Figure~\ref{fig:zamsplot}), and are less accurate (but still within $\sim 35-50\%$) for highly centrally concentrated and evolved stars, which have longer peak timescales and (correspondingly) smaller peak fallback rates than predicted. In contrast, the frozen-in approximation \citep{lkp} predicts peak timescales and peak fallback rates that are discrepant by $\gtrsim2$ orders of magnitude for the same stars (see Figure 1 of ~\citealt{bandopadhyay24}). We also provided a relativistic generalization of the MG model, obtained by replacing the Newtonian gravitational potential with the Paczy\'nski-Wiita potential.

The numerical simulations performed in this work are in good agreement with those of previous investigations of the complete disruption of main sequence stars by SMBHs. \cite{goicovic19} used the moving-mesh code {\sc arepo}~\citep{springel10} to study the disruption of a $1M_\odot$ ZAMS star by a $10^6M_\odot$ SMBH, and found a critical impact parameter $\beta_{\rm c}\sim1.8$, and a comparable estimate for $t_{\rm peak}$ and $\dot{M}_{\rm peak}$ to the ones shown here in Figure~\ref{fig:zamsplot}.~\cite{golightly19} simulated the disruption of three {\sc mesa} evolved stars, with masses $0.3 \, \Msun$, $1 \, \Msun$ and $3 \, \Msun$ at three different ages along the main sequence. They performed simulations for a single impact parameter $\beta=3$, and noted that the high mass and evolved stars (specifically the $1M_\odot$ TAMS star and the $3M_\odot$ MAMS and TAMS stars) were not completely destroyed at $\beta=3$, which is in agreement with the $t^{-9/4}$ scaling that we find here for the same stars (and at $\beta_{\rm c}>3$ for all three stars).

\cite{lawsmith20} performed simulations of the disruption of main sequence stars with masses ranging from $\sim 0.1-10 M_\odot$ by a $10^6M_\odot$ SMBH, using the Eulerian code {\sc flash} \citep{fryxell00}. The critical impact parameters at which they found complete disruption of ZAMS stars, 
as well as the peak timescale and peak magnitude of the corresponding fallback rates, are in good agreement with the results of the {\sc phantom} simulations presented here (and, hence, with the MG model). They noted that the highly centrally concentrated TAMS stars (specifically, the $1.0M_\odot$ and $1.5M_\odot$ TAMS stars) did not get destroyed at the maximum $\beta$ encounters simulated in their work, and extrapolated their scaling for the critical $\beta$ (based on the ratio of the central density to the average density, and obtained as an empirical fit to the results of the numerical simulations) to conclude that the critical impact parameter for complete disruption should be $\beta_{\rm c}=7.0$ and $\beta_{\rm c} =10.0$ for the $1.0M_\odot$ and $1.5M_\odot$ TAMS stars respectively (the highest impact parameters for their simulations being $\beta=6$ and $\beta=6.7$, for which, as shown in their Figure 7, the fallback rates exhibit the same scaling as the partial disruptions). For the $3M_\odot$ TAMS star, they find $\beta_{\rm c} =8.5$.

\cite{ryu20a} performed general relativistic hydrodynamic simulations of the disruption of {\sc mesa} evolved stars, with masses ranging from $0.15-10.0M_\odot$, for six SMBH masses between $10^5-5\times10^7M_\odot$. Since their simulations accounted for relativistic effects, they found smaller critical impact parameters for any given star compared to the predictions of the Newtonian MG model. Here, we provide a simple analytical extension of the MG model to account for relativistic effects, by using the Paczy{\'n}ski-Witta potential to model the tidal field of the SMBH. We find good agreement between the critical impact parameters estimated using this approach and those reported in numerical investigations of TDEs that include relativistic effects, either by using a generalized Newtonian potential (e.g.,~\citealt{gafton15,gafton19}) or by using relativistic hydrodynamics codes (e.g.,~\citealt{ryu20c,jankovic23}). Among these works, ~\cite{gafton15,gafton19} used a $\gamma=5/3$ polytropic structure to model the disrupted star, whereas \cite{ryu20c} and \cite{jankovic23} considered stellar structures obtained using {\sc mesa}. \cite{ryu20c} simulated the disruption of three stars ($M_\star=0.3,1.0,3.0 M_\odot$) at half their main-sequence lifetime, for a range of SMBH masses. \cite{jankovic23} simulated four stellar masses, $M_\star=0.6,1,2,3M_\odot$ at the ZAMS and TAMS stages, for $M_\bullet=10^6 M_\odot$ and a range of SMBH spins. We compared the predictions obtained using the Paczy{\'n}ski-Witta potential to their simulations for non-spinning black holes, and found good agreement in the crictical $\beta$ for complete disruption for the ZAMS stars. For the TAMS stars, the estimate for $\beta_{\rm c}$ obtained using the Paczy{\'n}ski-Witta potential is systematically smaller than the values found in \cite{jankovic23}. However, we note that for extreme parameters, where the stellar pericentre is $\lesssim 10$ gravitational radii (as is true for the TAMS stars), a more accurate treatment of relativistic gravity than the Paczy{\'n}ski-Witta potential is probably required. 

While we did not perform any simulations that account for relativistic effects in this work, it is likely that the fallback rates themselves are still fairly accurate, because the larger complete-disruption radius implies that the Newtonian approximation for the energy of the debris is more appropriate. The simulations performed in \cite{gafton19} demonstrate this directly for a sun-like $5/3-$polytropic star being disrupted by a $10^6 M_\odot$ SMBH -- Figure 9 in their paper shows that the time to peak and peak fallback rate are insensitive to relativistic effects near the complete disruption threshold, and in general differ by less than $\sim 10$ days. The simulations of \cite{jankovic23} consider stars over a wider range of mass and age and show more differences (at the  $\sim 20-30 \%$ level) between relativistic and non-relativistic treatments, but it is not clear how large these differences are near the complete disruption threshold (as their range in $\beta$ was fairly sparse). Those authors also note that different prescriptions for the self-gravitational field lead to qualitative and quantitative differences in the properties of the fallback rate, and hence it is difficult to definitively address the accuracy of the Newtonian approximation without a firm understanding of the relativistic case.

For stars near ZAMS, the model in excellent agreement with numerical simulations (as shown in Figure~\ref{fig:zamsplot}), while the agreement is less strong for more evolved and massive stars. TDEs from massive stars are likely to be rarer owing to their suppression in the IMF and comparatively short lifetimes, although massive-star disruptions could be more common in environments such as the Galactic Center, which is often modeled using top-heavy IMFs~\citep{smith01,bartko10} that lead to an over-representation of massive stars compared to a standard Kroupa IMF~\citep{kroupa01}.  Additionally, newly discovered ``extreme nuclear transients'' \citep{hinkle25,wiseman25}, which are among the most energetic transients and possess luminosities exceeding $\sim 2\times 10^{45} \, \mathrm{erg \, s^{-1}}$, are also consistent with the disruption of high-mass ($M_\star \gtrsim 3 M_\odot$) stars by high-mass ($M_\bullet \gtrsim 10^8 M_\odot$) SMBHs. Repeating partial tidal disruption events (a class of nuclear transients that repeat on timescales of months to years), and in particular those systems that have flared many times, such as ASASSN-14ko~\citep{payne21} and eRASSt-J045650 \citep{liu24}, are also most readily explained if the star being partially stripped is massive~\citep{liu23,bandopadhyay24b,liu25}.  Contrarily, low-mass and less evolved stars are unstable to mass transfer, and would be completely destroyed within $\sim1-2$ encounters following partial disruption~\citep{hjellming87,ge10,dai13,bandopadhyay25,yao25}. 

While these systems suggest that massive-star disruptions could be more common than might be expected from a typical IMF, the MG model is at most 30-50\% inaccurate compared to numerical simulations of such disruptions, whereas the frozen-in model is discrepant (with the same simulations) by orders of magnitude. The MG model can thus provide useful observational constraints on, e.g., the SMBH mass (which is most strongly correlated with the peak timescale compared to the stellar properties; \citealt{guillochon13, bandopadhyay24}) from observed TDE lightcurves.  Since the peak timescale in our model (and simulations) is measured relative to the pericenter passage of the star, the rise time for the lightcurve of an observed TDE will in general be shorter than the peak timescale predicted by the model. However, the sample of $\sim30$ spectroscopically classified TDEs detected by the Zwicky Transient facility~\citep{hammerstein23} all exhibit peak timescales of $\sim 30-50$ days, which lie in the range of $t_{\rm peak}$ values predicted for complete disruptions caused by SMBH masses $\sim 10^6 M_\odot$, thus upholding the predictions of the MG model. We have therefore demonstrated that the MG model provides a simple and accurate prediction of the peak fallback rate and the time at which this peak occurs, and we have extended this model to account for relativistic effects. 

\section*{Acknowledgements}
We thank the anonymous referee for useful comments and suggestions that improved the manuscript. J.F.~acknowledges support from Syracuse University through the SOURCE program and an Emerging Research Fellows award. A.B.~acknowledges support from NASA through the FINESST program, grant 80NSSC24K1548. E.R.C.~acknowledges support from NASA through the Astrophysics Theory Program, grant 80NSSC24K0897. Additional support for this work (A.B.) was provided by the National Aeronautics and Space Administration through Chandra Award Number 25700383 issued by the Chandra X-ray Observatory Center, which is operated by the Smithsonian Astrophysical Observatory for and on behalf of the National Aeronautics Space Administration under contract NAS8-03060. C.J.N.~acknowledges support from the Science and Technology Facilities Council (grant No. ST/Y000544/1) and from the Leverhulme Trust (grant No. RPG-2021-380).
\clearpage

\appendix
\section{Convergence of Numerical Simulations}
\label{sec:convergence}
\begin{figure*}[h!]
    \includegraphics[width=0.51\textwidth]{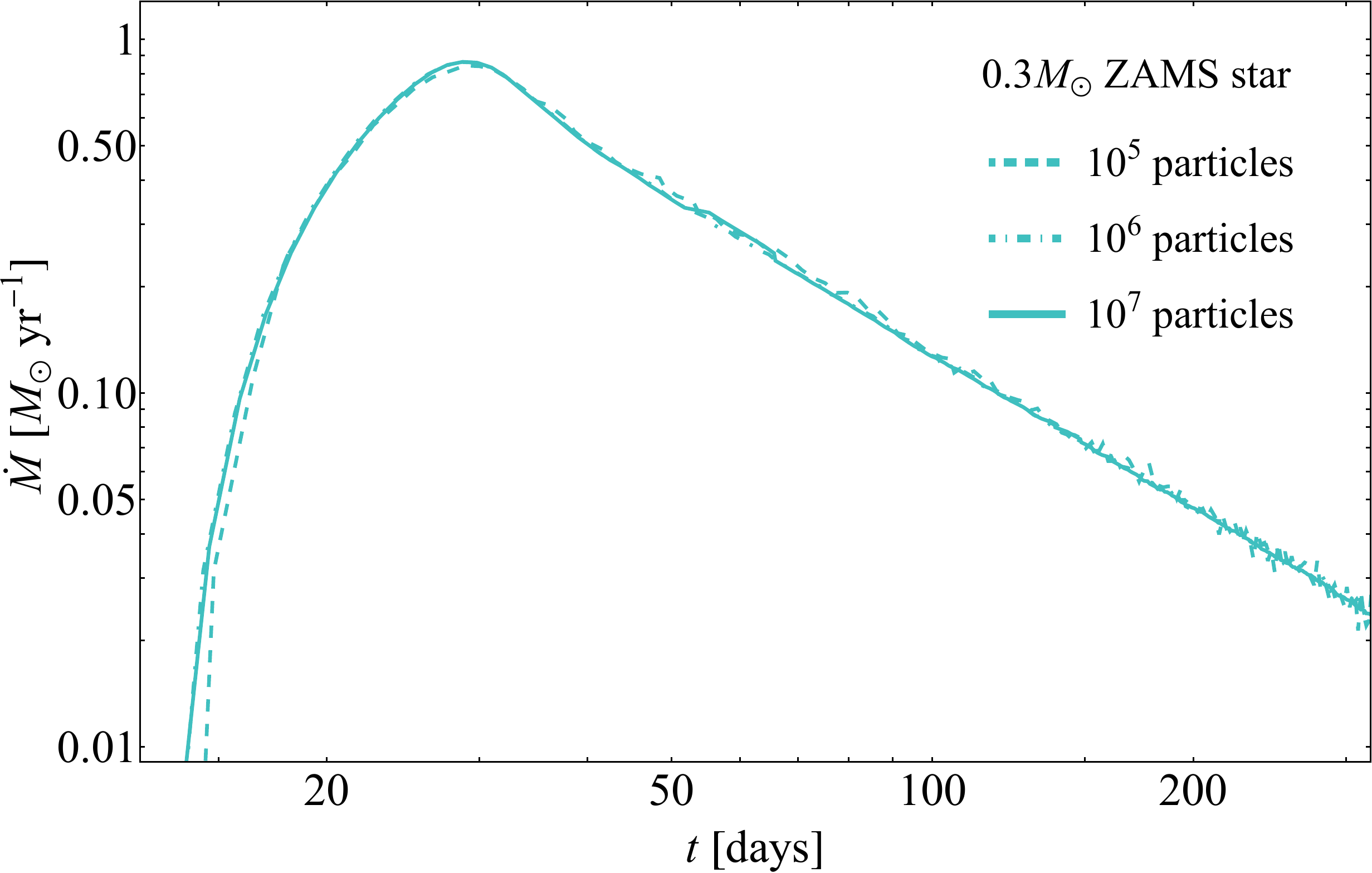} 
    \includegraphics[width=0.51\textwidth]{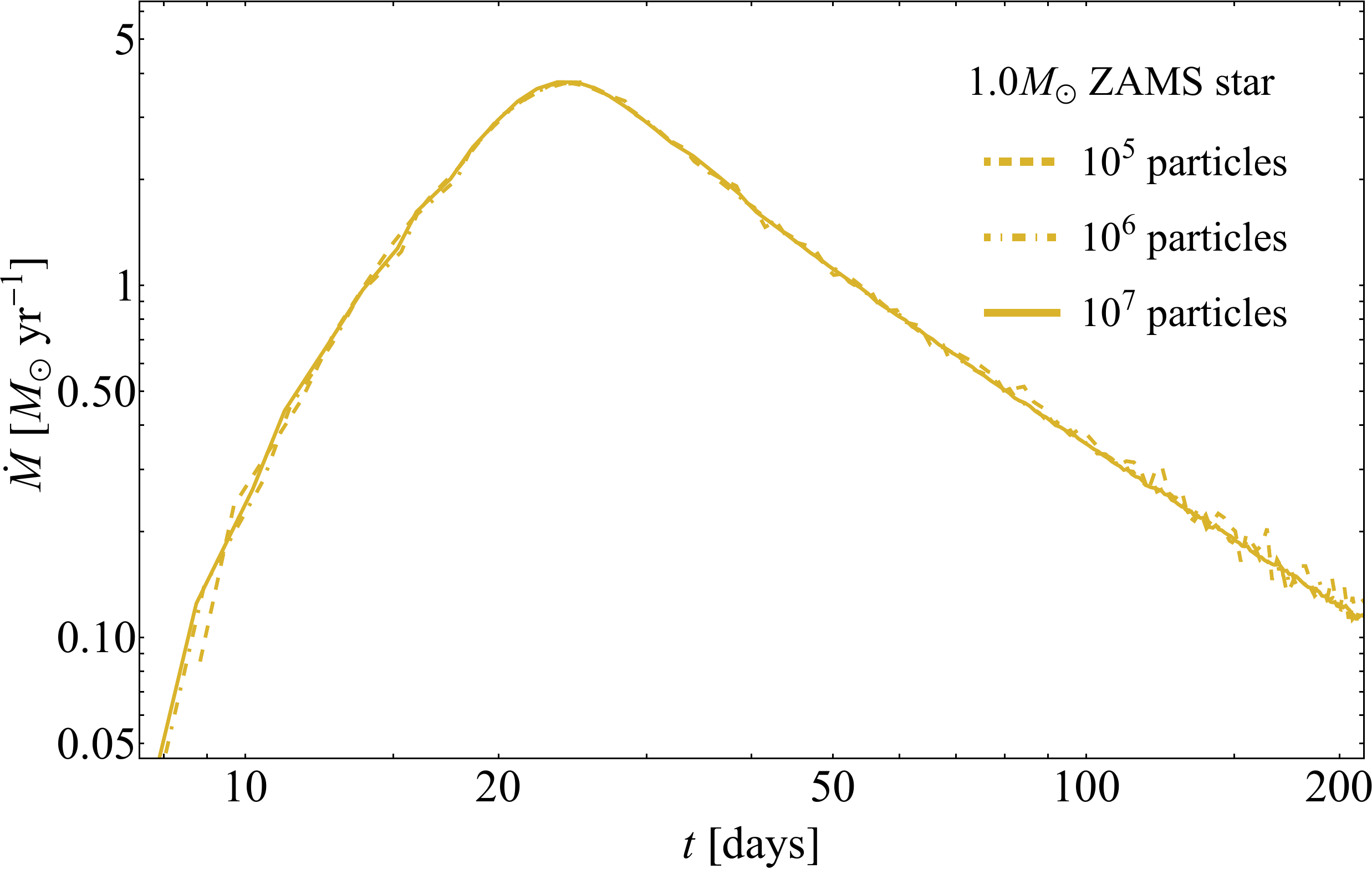} \\
    \includegraphics[width=0.51\textwidth]{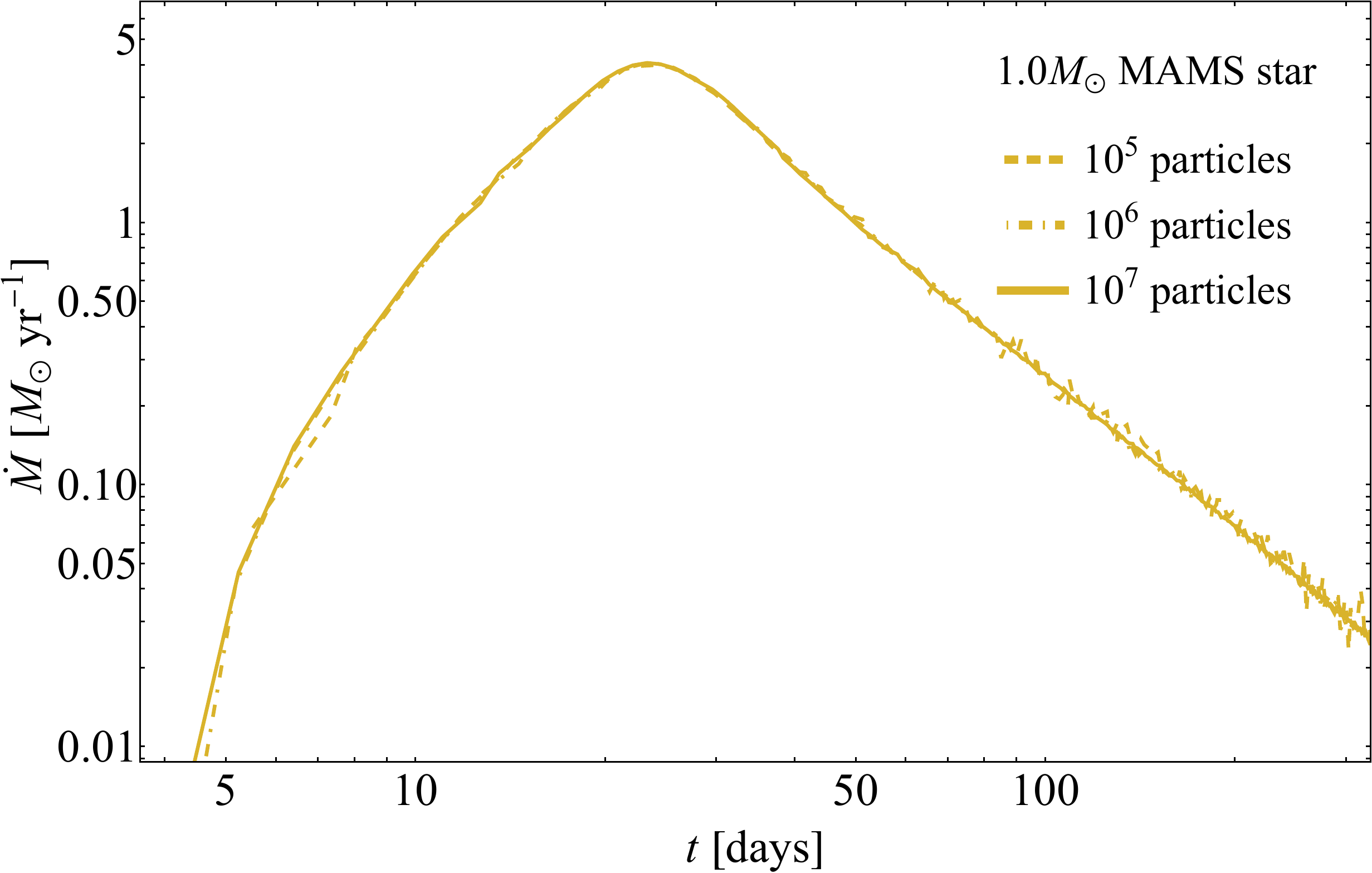} 
    \includegraphics[width=0.51\textwidth]{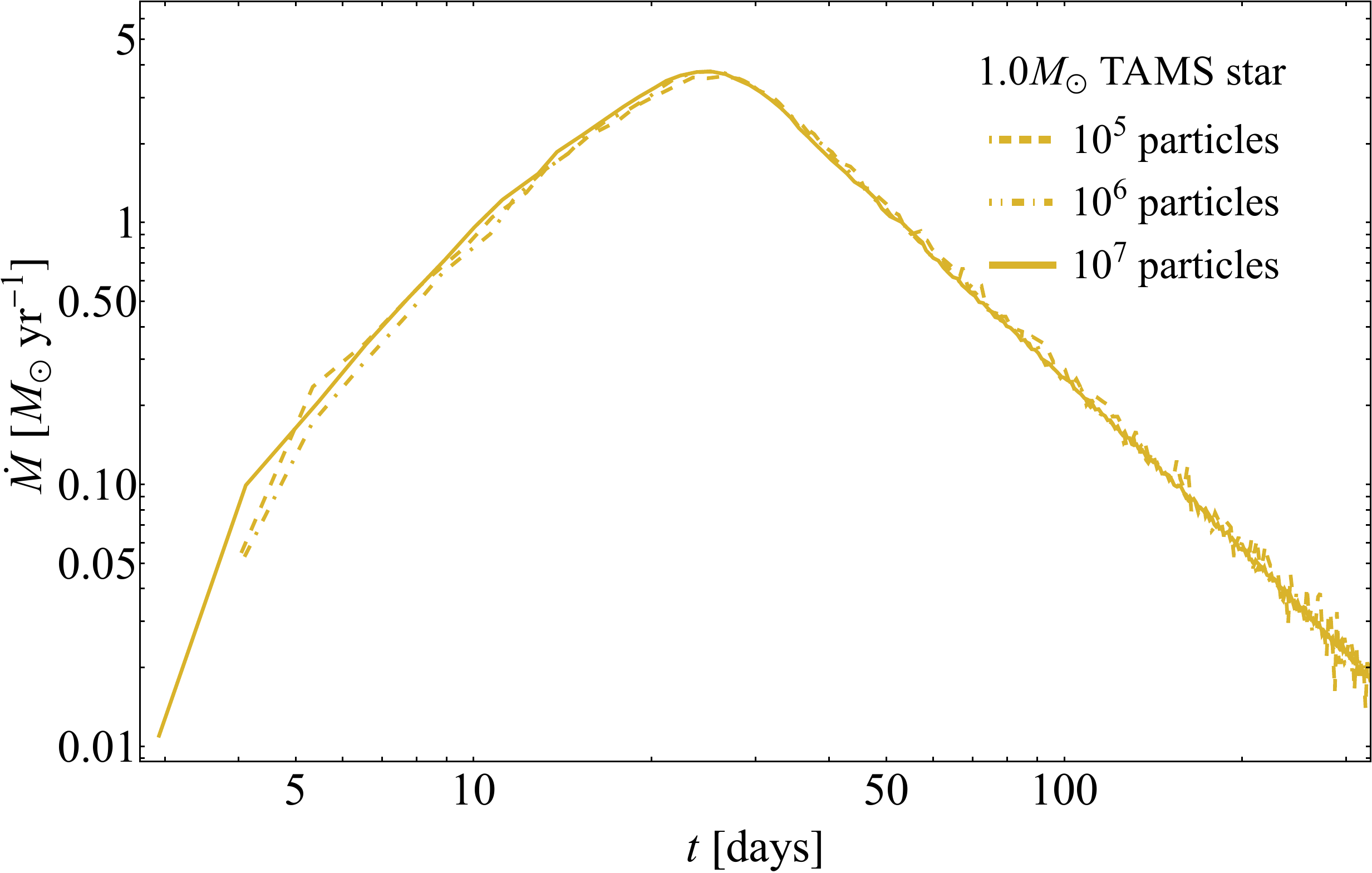} 
    \caption{Fallback rates at three different numerical resolutions from the disruption of a $0.3 M_\odot$ ZAMS star ($\beta_{\rm c} =0.96$; top-left), $1.0M_\odot$ ZAMS star ($\beta_{\rm c} =1.80$; top-right), $1.0M_\odot$ MAMS star ($\beta_{\rm c} =2.58$; bottom-left) and $1.0M_\odot$ TAMS star ($\beta_{\rm c} =4.10$; bottom-right), on a $\beta=\beta_{\rm c}$ (as predicted by the MG model) parabolic orbit about a $10^6M_\odot$ SMBH.} 
    \label{fig:convergence-tests}
\end{figure*}

To assess the convergence of our simulations, we performed a subset of the simulations presented in Section~\ref{sec:MG-model} at $10^5$ and $10^7$ particles, compared to the fiducial resolution of $10^6$ particles used for the results presented in Section~\ref{sec:MG-model}. Figure~\ref{fig:convergence-tests} shows the fallback rates obtained using the three different resolutions for a $0.3 M_\odot$ ZAMS star (top-left), $1.0M_\odot$ ZAMS star (top-right), $1.0M_\odot$ MAMS star (bottom-left) and $1.0M_\odot$ TAMS star (bottom-right). The simulation setup is identical to the description provided in Section~\ref{sec:MG-model}. As seen in the figure, the peak of the fallback rate is almost identical at the three different resolutions, demonstrating the numerical convergence of the results shown in Figures~\ref{fig:zamsplot}-\ref{fig:mams+tamsplot}.

\bibliographystyle{aasjournal}

\end{document}